\documentclass[journal,onecolumn,11pt]{IEEEtran}

\RequirePackage{times}
\RequirePackage[usenames]{pstricks} 
\RequirePackage{rotating}
\RequirePackage{indentfirst} 
\RequirePackage{color} 
\RequirePackage{graphicx} 
\RequirePackage{bmpsize} 
\RequirePackage{epsfig} 
\RequirePackage{eepic} 
\RequirePackage{subfigure} 
\RequirePackage{wrapfig} 
\RequirePackage{picinpar} 
\RequirePackage{amsmath,amssymb,amsthm,amsbsy,amsfonts,dotlessj} 
\RequirePackage{latexsym}
\RequirePackage{mathptm} 
\RequirePackage{mathptmx} 
\DeclareMathVersion{bold} 
\RequirePackage{bm} 
\RequirePackage{multicol}
\RequirePackage{esvect} 
\RequirePackage{ifthen}
\RequirePackage{etoolbox}
\RequirePackage{relsize}

\RequirePackage{setspace}
\RequirePackage{moreverb}
\RequirePackage{ulem} 
        \let\em=\it 
\RequirePackage[sectionbib]{chapterbib}
\RequirePackage{fancyhdr}
\RequirePackage{array}
\RequirePackage{floatflt} 
\RequirePackage{srcltx}
\RequirePackage{chemarrow}
\RequirePackage{fixme}
\RequirePackage[footnotesize,bf,figurename=Fig.,hang]{caption}
\RequirePackage{hyperref} 

\usepackage{./tweaklist} 

\providecommand{\subparagraph}{} 
\RequirePackage[small,compact]{titlesec} 
\RequirePackage[square,sort&compress,numbers]{natbib} 
\RequirePackage{lipsum}
\RequirePackage{microtype}
\RequirePackage{url}
\RequirePackage{listings} 
\RequirePackage{algorithmic}
\RequirePackage[boxed,algoruled,linesnumbered,vlined,nofillcomment]{algorithm2e}
\RequirePackage{multirow}
\RequirePackage{booktabs}

\fxsetup{ status=draft,
	  	  layout=footnote,
		  layout=inline, 
		  theme=color,
	  	  marginclue
}
\definecolor{mydarkgreen}{rgb}{0.0,0.5,0.0}
\definecolor{myred}{rgb}{0.9,0.0,0.0}

\newcommand{\matlabscript}[2]{
    \begin{itemize}
    \item[]\lstinputlisting[caption=#2,label=#1]{#1}
    \end{itemize}
}

\definecolor{mylightgrey}{rgb}{0.6,0.6,0.6}
\definecolor{mydarkgrey}{rgb}{0.3,0.3,0.3}
\definecolor{mybrown}{rgb}{0.6,0.2,0}
\definecolor{mylightgreen}{rgb}{0,1,0}
\definecolor{mydarkgreen}{rgb}{0,0.5,0}
\definecolor{MyDarkGreen}{rgb}{0.0,0.4,0.0} 
\definecolor{myorange}{rgb}{1,0.4,0}
\definecolor{mypurple}{rgb}{0.8,0,0.8}
\definecolor{MyDarkRed}{rgb}{0.4,0.0,0.0} 

\newcommand{\ignore}[1]{}

\newcommand{\COMMENT}[1]{}

\newcommand{\redHL}[1]{{\color{red}\bf #1}}

\newcommand{\jnote}[1]{{\color{mydarkgreen}\textit{\uline{#1}}}}

\newcommand{\eg}{{\it e.g.\/}}
\newcommand{\ie}{{\it i.e.\/}}
\newcommand{\wrt}{{\it wrt}}

\newcommand{\aka}{{\it aka.\/}}
\ifx \etc \undefined                                                             
  \def\etc#1{\textit{etc}\if#1..\else.#1\fi}                                     
\fi

\newcommand{\eqnref}[1]{(\ref{eq:#1})}
\newcommand{\eqnlabel}[1]{\label{eq:#1}}
\newcommand{\figlabel}[1]{\label{fig:#1}}
\newcommand{\figref}[1]{Fig.~\ref{fig:#1}}

\newcommand{\seclabel}[1]{\label{sec:#1}}
\newcommand{\secref}[1]{Sec.~\ref{sec:#1}}

\newcommand{\applabel}[1]{\label{app:#1}}
\newcommand{\appref}[1]{Appendix~\ref{app:#1}}
\newcommand{\tablabel}[1]{\label{tab:#1}}
\newcommand{\tabref}[1]{Table~\ref{tab:#1}}

\newcommand{\thmlabel}[1]{\label{thm:#1}}

\newcommand{\lemlabel}[1]{\label{lem:#1}}

\newcommand{\httpref}[1]{}

\renewcommand{\jmath}{j}

\newcommand{\halmos}{\hskip\textwidth minus\textwidth \rule{6pt}{6pt}}

\newcommand{\be}[1]{\begin{equation}\eqnlabel{#1}}
\def\ee{\end{equation}}

\newtheorem{lemma}{Lemma}[section]

\newtheorem{theorem}{Theorem}

\newcommand{\numberedtheorem}[3]{
	\ifthen{\not\equal{#1}{}}{
		\let\oof=\thetheorem
		\renewcommand{\thetheorem}{#1}
	}
	\begin{theorem}
	\ifthen{\not\equal{#2}{}}{
		\thmlabel{#2}
	}
	#3
	\end{theorem}
	\ifthen{\not\equal{#1}{}}{
		\let\thetheorem=\oof
		\addtocounter{theorem}{-1}
	}
}
\newcommand{\numberedlemma}[3]{
	\ifthen{\not\equal{#1}{}}{
		\let\oof=\thelemma
		\renewcommand{\thelemma}{#1}
	}
	\begin{lemma}
	\ifthen{\not\equal{#2}{}}{
		\lemlabel{#2}
	}
	#3
	\end{lemma}
	\ifthen{\not\equal{#1}{}}{
		\let\thelemma=\oof
		\addtocounter{lemma}{-1}
	}
}

\newcommand{\modifiedListParamters}{
  \setlength{\topsep}{0pt}
  \setlength{\itemsep}{-0.1em}
  \setlength{\partopsep}{0em}
  \setlength{\parsep}{0em}
  \setlength{\leftmargin}{1.1em}
  \setlength{\itemindent}{0.0em}
}
\renewcommand{\enumhook}{
  \modifiedListParamters
  \setlength{\itemindent}{0.2em}
}
\renewcommand{\itemhook}{
  \modifiedListParamters
}
\renewcommand{\descripthook}{
  \modifiedListParamters
  \setlength{\itemindent}{-1.0em}
}

	\titlespacing{\section}{0pt}{*0}{*0.1} 
	\titlespacing{\subsection}{0pt}{*0.1}{*0}
	\titlespacing{\subsubsection}{0pt}{*0.1}{*0}

\makeatletter
\g@addto@macro\normalsize{
  \setlength\abovedisplayskip{1pt}
  \setlength\belowdisplayskip{1pt}
  \setlength\abovedisplayshortskip{1pt}
  \setlength\belowdisplayshortskip{1pt}
}
\makeatother

\def\ssp{\def\baselinestretch{0.9}\large\normalsize} 
 
\ssp

\RequirePackage[top=3.0cm,bottom=3.0cm,left=2.5cm,right=2.5cm,nohead,nofoot]{geometry}

\newcommand{\MATLAB}{MATLAB$^\text{\textregistered}$}
\newcommand{\Spectre}{Spectre$^\text{\textregistered}$}
\newcommand{\HSPICE}{HSPICE}

\definecolor{MyDarkGreen}{rgb}{0.0,0.4,0.0} 
\definecolor{MyDarkRed}{rgb}{0.4,0.0,0.0} 
\renewcommand{\matlabscript}[3]{
    \begin{itemize}
    \item[]\lstinputlisting[language=Matlab,caption=#3,label=#2]{#1}
    \end{itemize}
}

\newcommand{\verilogascript}[3]{
    \begin{itemize}
    \item[]\lstinputlisting[language=C,caption=#3,label=#2]{#1}
    \end{itemize}
}

\begin{document}

\renewcommand{\refname}{References}

\let\redHL=\ignore
\let\jnote=\ignore
\let\tnote=\ignore
\let\tsumm=\ignore

\pagestyle{empty}

\title{\Huge{Well-Posed Models of Memristive Devices}}
\author{Tianshi Wang and Jaijeet Roychowdhury \vspace{0.35em} \\
\small{Department of Electrical Engineering and Computer Sciences, University 
                            of California, Berkeley, CA, USA} \\
\normalsize{email: {$\texttt{\small \{tianshi, jr\}@berkeley.edu}$}}}

\maketitle

\begin{abstract}
Existing compact models for memristive devices (including RRAM and CBRAM) all suffer from issues related to mathematical ill-posedness and/or improper implementation. 
This limits their value for simulation and design and in some cases, results in qualitatively unphysical predictions.
We identify the causes of ill-posedness in these models.
We then show how memristive devices in general can be modelled using only continuous/smooth primitives in such a way that they always respect physical bounds for filament length and also feature well-defined and correct DC behaviour.
We show how to express these models properly in languages like Verilog-A and ModSpec (\MATLAB{}).
We apply these methods to correct previously published RRAM and memristor models and make them well posed. 
The result is a collection of memristor models that may be dubbed ``simulation-ready'', \ie, that feature the right physical characteristics and are suitable for robust and consistent simulation in DC, AC, transient, \etc, analyses.
We provide implementations of these models in both ModSpec/\MATLAB{} and Verilog-A.

\end{abstract}

\section{\normalfont {\large Introduction}}
\thispagestyle{empty}

In 1971, Leon Chua noted \cite{chua1971memristor} that while two-terminal circuit elements relating voltage and current (\ie, resistors), voltage and charge (capacitors) and current and flux (inductors) were well known, no element that directly relates charge and flux seemed to exist. 
He explored the properties of this hypothetical element and found that its voltage-current characteristics would be those of a resistor, but that if the element were nonlinear, its resistance would change with time and be determined by the history of biasses applied to the device. 
In other words, the instantaneous resistance of the element would retain some memory of past inputs. 
Chua dubbed this missing element a ``memristor'', and showed that a telltale characteristic was that its $i$--$v$ curves would always pass through $(0,0)$, regardless of how it was biassed as a function of time.\footnote{\ie, a memristor's $i$--$v$ characteristics are ``pinched'' at the origin.}
Long after Chua's landmark observation, devices with memristive behaviour were found in nature, \eg, in the well-publicized nano-crossbar device of Stan Williams and colleagues \cite{williamsIEEEspectrum2008,StSnDuWi2008memristorFound}, and others as well \cite{JoLu2009crossbar,samsung2009crossbar}. 
It was also realized that many physically observed devices prior to \cite{williamsIEEEspectrum2008,StSnDuWi2008memristorFound} were in fact memristors \cite{thakoor1990solid,buot1994binary,erokhin2008memristor}.

Physically, present-day memristive nano-devices typically operate by forming and destroying conducting filaments through an insulating material sandwiched between two contacts separated by a small distance $l$.  
The conducting filaments can be of different types. 
For example, they can consist of oxide vacancies, by filling which electrons can flow, as in RRAM (Resistive Random Access Memory \cite{wong2012RRAM}). 
In CBRAM (Conductive Bridging RAM \cite{kund2005CBRAM}), metal ions\footnote{various metals, including Cu, Ag, W, Sn and Cr, have been used.} that infiltrate the insulator form the conducting filament. 
In memristors made of Si-impregnated silica \cite{mehonic2012siliconRRAM}, conduction occurs via tunnelling between traps.
Depending on the magnitude and polarity of the voltage applied, the conducting filaments can lengthen or shorten; it is their length that determines the resistance of the device. 
Basic geometry indicates that the length of the filament(s) must always be between zero (\ie, there is no filament) and the distance between the contacts (\ie, the filament connects the two contacts) --- in other words, the length of the filament(s) must never be outside the range $[0,l]$. 
Another basic property is that when the voltage across the device has one polarity (say positive), the filament grows until it reaches its maximum length $l$, at which it settles; whereas for the opposite polarity (say negative), the filament shrinks until it reaches its minimum length $0$. Therefore, if a positive DC voltage is applied, the DC (\ie, long term) response of the memristor's filament length must be $l$; whereas if a negative DC voltage is applied, its DC response must be $0$.

A number of novel circuits based on memristors have been proposed \cite{YaStSt2013MemristorComputing,vourkas2015memristor}, most of which use crossbar architectures for non-volatile memory \cite{LiRoKuWa2010complementary,ShChLi2009fastRRAM} and neuromorphic computing \cite{AlZaSt2013NNcrossbar,ShFaAiRo2014nonBooleanRRAM,JoChLu2010memsynapse} applications. 
To support their design, various compact models of memristors, purportedly suitable for simulation in SPICE-like simulators, have been published.
However, our attempts to use these models have revealed shortcomings serious enough to preclude their general use for simulation or design. 
Broadly speaking, these existing models suffer from \textit{ill-posedness} issues; \eg, they are not properly defined at all biasses, or their outputs are not unique, or they suffer from continuity/smoothness problems.  
Well-posedness \cite{hadamard1902problemes, parker1989mcgraw}\footnote{A well-posed mathematical model of a physical device should have the properties that a unique solution or output should exist for any given input, and that outputs should vary smoothly with respect to inputs and parameters.} is a fundamental requirement for models meant to represent physical reality and is also crucial for numerical algorithms using the models to work properly.

The well-posedness requirement applies not only to memristive devices, but to any model meant for simulation. 
To appreciate why, it is important to realize that a model represents a \textit{mathematical abstraction} of a physical device.  
While this abstraction must represent reality well enough to be useful for prediction, it must also be suitable for use with numerical simulation algorithms. 
To be so, it needs to satisfy certain important mathematical properties, the most basic and universal of which is well posedness.

To illustrate how a well-posed mathematical model must often be ``more than'' the physical device it represents, consider the question: is it necessary to model a device outside regions that are physically reasonable in proper operation?\footnote{This is a frequent point of confusion amongst compact model developers.}
For example, should a compact model of a memristor (or a diode, or resistor, or IC MOSFET), be ``valid'' at a bias of a million volts (at which, in reality, most physical devices would simply burn up)?  
The answer to this question is yes -- indeed, it has been a standard requirement for device models (including resistors, capacitors, diodes, BJT and MOS devices, \etc) in SPICE-like simulators to evaluate successfully and provide unique, smoothly varying outputs at all biasses, including large, physically unrealistic, biasses. 
These requirements stem not only from the numerical algorithms used by simulators (in particular, the Newton-Raphson (NR) algorithm for solving nonlinear equations \cite{numericalrecipes}), but also from the iterative methodology using which circuits are typically designed:
\begin{enumerate}
\item In the process of converging to a valid solution of the circuit, NR typically applies a sequence of biasses to devices; many of these biasses can be large or physically unreasonable. 
  Compact models must be designed to evaluate successfully and be smooth at \textit{every} bias applied, whether it is physically reasonable or not, in order for NR to go about trying to find a solution \cite{RoNOWmonograph2009, WaRo2013VAnanoHUB}.  
  If devices are modelled well and the circuit has been designed properly, then, \textit{at the solution found by NR}, biasses to devices can be expected to be physically reasonable.

\item Even if NR converges to a solution that is physically unreasonable, the ``bad'' solution has value in circuit design, for it  typically provides quantitative insight into what is wrong with the design.
  A compact model that refuses to evaluate or generates a floating-point error prevents such solutions, and the insights they provide, from being found.
\end{enumerate}

Common ill-posedness mechanisms in models include division-by-zero errors, often due to expressions like $\frac{1}{x-a}$, which become unbounded (and feature a ``doubly infinite'' jump) at $x\!=\!a$; the use of $\log()$ or $\sqrt()$ without ensuring that their arguments are always positive, regardless of bias; the fundamental misconception that non-real (\ie, complex) numbers or infinity are legal values for device models (they are not!); and ``sharp''/``pointy'' functions like $|x|$, whose derivatives are not continuous.

Another key aspect of well-posedness is that the model's equations must produce mathematically valid outputs for any mathematically valid input to the device.
Possibly the most basic kind of input is one that is constant (``DC'') for all time. 
DC solutions (``operating points'') are fundamental in circuit design; they are typically the kind of solution a designer seeks first of all, and are used as starting points for other analyses like transient, small signal AC, \etc. 
If a model's equations do not produce valid DC outputs given DC inputs, it fails a very fundamental well-posedness requirement. 
For example, the equation $\frac{d}{dt}o(t) = i(t)$ is ill posed, since no DC (constant) solution for the output $o(t)$ is possible if the input $i(t)$ is any non-zero constant. 
Such ill posedness is typically indicative of some fundamental physical principle being violated; for example, in the case of $\frac{d}{dt}o(t) = i(t)$, the system is not strictly stable \cite{zadeh1}.
Indeed, a well-posed model that is properly written and implemented should work consistently in every analysis (including DC, transient, AC, \etc.).\footnote{Different analyses correspond to different ways of exciting the device, or to finding specific kinds of outputs. For example, periodic steady state analyses  excite the device with time-periodic inputs, and seek similarly time-periodic outputs.}

In spite of seemingly significant efforts to devise memristor models, \textit{every} model we are aware of\footnote{We request the reader to contact us if he/she is aware of published, openly available and reproducible memristor models prior to this work (other than the ones noted here), especially if they do not suffer from ill-posedness issues.} in the literature \cite{UMich2011RRAM,Stanford2012RRAMspice,Stanford2014RRAMverilog,li2015RRAM,ASU2015RRAMverilog,xu2014twoMemristors} suffers from one or more of the above-mentioned types of ill-posedness.
The University of Michigan model \cite{UMich2011RRAM}, many aspects of which have been adopted by later models, suffers from division-by-zero errors and DC response problems.
An RRAM model from Stanford/ASU with several variants \cite{Stanford2012RRAMspice,Stanford2014RRAMverilog,li2015RRAM,ASU2015RRAMverilog,kvatinsky2011memristor,yakopcic2013memristor,yakopcic2014memristor,kvatinsky2013team} that has received considerable publicity suffers from egregious DC response problems.\footnote{To their credit, the authors of \cite{Stanford2014RRAMverilog} explicitly note this deficiency in their model's documentation.} 
The UESTC memristor models \cite{xu2014twoMemristors}, though they avoid many issues common to other models, still suffer from subtle (but serious) DC response problems.
The TEAM models for general memristors \cite{kvatinsky2011memristor,yakopcic2013memristor,yakopcic2014memristor,kvatinsky2013team} also suffer from DC, uniqueness and continuity/smoothness issues. 
Over and above well-posedness issues, released versions of existing memristor compact models frequently suffer from deficiencies in the way their equations are expressed in modelling languages like Verilog-A. 
Examples of deficiencies we have encountered include attempts to perform time-integration of differential equations within the model definition, inserting time-varying noise terms as an integral part of the model, using integral formulations instead of differential ones, \etc. 
As explained in \cite{WaRo2013VAnanoHUB}, such practices compromise accuracy, limit the model's ability to support all analyses, reduce portability across simulators, and so on.
Further details about the shortcomings we have observed in these models are provided later in this paper.

In this paper, we explain the correct generic way to model memristive devices in a well-posed manner. 
Our modelling technique sets up the dynamics of the filament length using a differential equation, and the current-voltage relationship of the memristor using an algebraic equation involving the filament length.
Employing only continuous/smooth mathematical constructs, we show how filament dynamics can be modelled such that physical bounds are always respected and correct DC behaviour for positive and negative biasses always results.
In the process of developing our modelling technique, we pinpoint several common mechanisms underlying ill-posedness in prior models.
Since filament dynamics in memristive devices have features closely related to hysteresis,\footnote{Memristive devices may be said to be at the ``cusp of hysteresis''.} we explain how to model hysteresis correctly in general, then apply this to memristive devices.  
We use our techniques to correct several previous models, making them well posed and suitable for any analysis (including DC, transient, AC, periodic steady state, \etc).
The process of restoring well-posedness to memristor models also provides insights into possible physical mechanisms in memristors that seem not to have been looked into yet.

The remainder of the paper is organized as follows. In \secref{hys}, we explain how hysteresis should be modelled in general, \ie, using internal unknown variables and implicit equations.
We illustrate how a model with internal unknowns and implicit equations can be written properly in both Verilog-A \cite{lemaitre2003extensions,coram2004VA,WaRo2013VAnanoHUB,mcandrew2015VA} and ModSpec \cite{amsallem2011modspec,WaAaWuYaRoCICC2015MAPP}.
In \secref{RRAM}, we specialize our general model template for hysteresis to RRAM devices, showing how to design the continuous/smooth equations involved so that filament length boundaries are always respected and the correct DC behaviour results.
We write well-posed RRAM models in both Verilog-A and ModSpec and test them in simulation, using DC, transient and homotopy \cite{RoMeTCAD2006} analyses.\footnote{Homotopy analysis provides considerable insight into RRAM behaviour.}
Then, in \secref{convergence}, we develop techniques for aiding numerical convergence in the RRAM model.
In particular, we design a SPICE-compatible limiting function for the rapidly-growing hyperbolic sine function used in the RRAM model, inspired by the limiting functions used in SPICE's non-linear semiconductor devices.
To our knowledge, this is the first limiting function designed to aid convergence after PNJLIM and FETLIM, which were developed as part of the original Berkeley SPICE \cite{Nagel1975}.
Next, in \secref{memristor}, we study all the published compact models for memristors we are aware of that come with concrete equations or code.  
We identify issues of ill-posedness and poor implementation that affect their applicability in simulation.
We then use our modelling techniques to correct their problems and turn them into well-posed models, providing proper implementations in ModSpec and Verilog-A.

The result of our study is a collection of well-posed, properly implemented, compact models for memristive devices.  
Specifically, we devise 5 different algebraic current-voltage and 6 different differential equation dynamical models for filament length, \ie, 30 different models for memristors and/or RRAM devices, all well posed.

Although we use underlying equations published by others, we modify them to remove ill-posedness issues, and also provide proper implementations.
Understanding the process by which we do this can be valuable for the development of future models, not only of memristors, but of other hysteretic devices as well.

\section{\normalfont {\large How to Model Hysteresis Properly}}\seclabel{hys}
To develop our memristor models, we first study how to model $i$--$v$ hysteresis in
two-terminal devices properly.
We show that the $i$--$v$ hysteresis can be modelled with the help of an internal
state variable and an implicit differential equation.
Then with the help of an example, we illustrate how a model with internal
unknowns and implicit equations can be properly written in both Verilog-A and
the ModSpec format.

\subsection{Model Template for Devices with $i$--$v$ Hysteresis}\seclabel{hys_eqn}

The equation of a general two-terminal resistive device can be written as
\begin{equation}\eqnlabel{IV_f}
	i(t) = f(v(t)),
\end{equation}
where $v(t)$ is the voltage across the device, $i(t)$ the current through it.
For example, the function $f(.)$ for a simple linear resistor can be written as
\begin{equation}\eqnlabel{R_f}
	f(v(t)) = \frac{v(t)}{R}.
\end{equation}

For devices with $i$--$v$ hysteresis, $i(t)$ and $v(t)$ cannot have a simple
algebraic mapping like \eqnref{IV_f}.
Instead, we introduce a state variable $s(t)$ into \eqnref{IV_f} and rewrite
the $i$--$v$ relationship as 
\begin{equation}\eqnlabel{IV_f1}
	i(t) = f_1(v(t),~s(t)).
\end{equation}

The dynamics of the internal state variable $s(t)$ is governed by a
differential equation:
\begin{equation}\eqnlabel{sV_f2}
	\frac{d}{dt}s(t) = f_2(v(t),~s(t)).
\end{equation}

The internal state variable $s(t)$ can have several physical meanings.
If we consider the original memristor model proposed by Chua in the 1970s
\cite{chua1971memristor,chua1976memristor}, $s(t)$ can be thought of as the
flux or charge stored in the device.
In the context of metal-insulator-metal-(MIM)-structured RAM devices, \eg,
RRAMs and CBRAMs, $s(t)$ can represent either the length of the conductive
filament/bridge, or the gap between the tip of the filament/bridge to the
opposing electrode.\footnote{For CBRAM devices, the tunnelling gap can also form
in the middle of the conductive bridge instead of on one of its ends
\cite{lin2012IEDM}.} 

In all these scenarios, $s(t)$ has some influence on $i(t)$.
So we cannot directly calculate the current based on the voltage applied to the
device at a single time $t$; $i(t)$ also depends on the value of $s(t)$.
On the other hand, at time $t$, the value of $s(t)$ is determined by the
history of $v(t)$ according to \eqnref{sV_f2}.
Therefore, we can think of the device as having internal ``memory'' of the
history of its input voltage.
If we choose the formula for $f_1$ and $f_2$ in \eqnref{IV_f1} and
\eqnref{sV_f2} properly, as we sweep the voltage, hysteresis in the current
becomes possible.

In the rest of this paper, \eqnref{IV_f1} and \eqnref{sV_f2} serve as a model
template for two-terminal devices with $i$--$v$ hysteresis.
To illustrate its use, we design a device example, namely
``\texttt{hys\_example}'', with functions $f_1$ and $f_2$ defined as follows.

\begin{equation}\eqnlabel{hys_f1}
	f_1(v(t), s(t)) = \frac{v(t)}{R} \cdot (\tanh(s(t)) + 1).
\end{equation}

\begin{equation}\eqnlabel{hys_f2}
	f_2(v(t), s(t)) = \frac{1}{\tau} \left( v(t) - s^3(t) + s(t) \right).
\end{equation}

The choice of $f_1$ is easy to understand.
$\tanh()$ is a monotonically increasing function with range $(-1, 1)$.
We add $1$ to it to make its range positive.
We then incorporate it into $f_1$ as a factor such that $s(t)$ can modulate the
conductance of the device between $0$ and $2/R$.

The choice of $f_2$ determines the dynamics of $s(t)$.
And when $f_2 = 0$, the corresponding ($v$, $s$) pairs will show up as part of
the DC solutions of circuits containing this device.
Therefore, if we plot the values of $f_2$ in a contour plot, such as in
\figref{hys_f2} (a), the curve representing $f_2 = 0$ is especially important.
Through the use of a simple cubic polynomial of $s(t)$ in \eqnref{hys_f2}, we
design the $f_2 = 0$ curve to fold back in the middle, crossing the $v = 0$
axis three times.
In this way, when $v$ is around $0$, there are three possible values $s$ can
settle on, all satisfying $\frac{d}{dt} s(t) = f_2 = 0$.
This multiple stability in state variable $s$ is the foundation of hysteresis
found in the DC sweep on the device.

\figref{hys_f2} (b) illustrates how hysteresis takes place in DC sweeps.
In \figref{hys_f2} (b), we divide the $f_2 = 0$ curve into three parts: curve
\texttt{A} and \texttt{B} have positive slopes while \texttt{C} has a negative
one.
When we sweep $v$ towards the right at a very slow speed to approximate DC
conditions, starting from a negative value left of $V-$, at the beginning,
there is only one possible DC solution of $s$. 
As we increase $v$, the ($v$, $s$) pair will move along curve \texttt{A}, until
\texttt{A} ends when $v$ reaches $V+$.
If $v$ increases slightly beyond $V+$, multiple stability in $s$ disappears.
($v$, $s$) reaches the $f_2 > 0$ region and $s$ will grow until it reaches
the \texttt{B} part of the $f_2 = 0$ curve.
This shows up in the DC solutions as a sudden jump of $s$ towards curve
\texttt{B}.
Similarly, when we sweep $v$ in the other direction starting from the right of
$V+$, the ($v$, $s$) pair will follow curve \texttt{B}, then have a sudden
shift to \texttt{A} at $V-$.
Because $V+ > V-$, hysteresis occurs in $s$ when sweeping $v$, as illustrated
in \figref{hys_f2} (b).
Since $s$ modulates the device's conductance, there will also be hysteresis in
the $i$--$v$ relationship.

\begin{figure}[htbp]
\centering{
    \epsfig{file=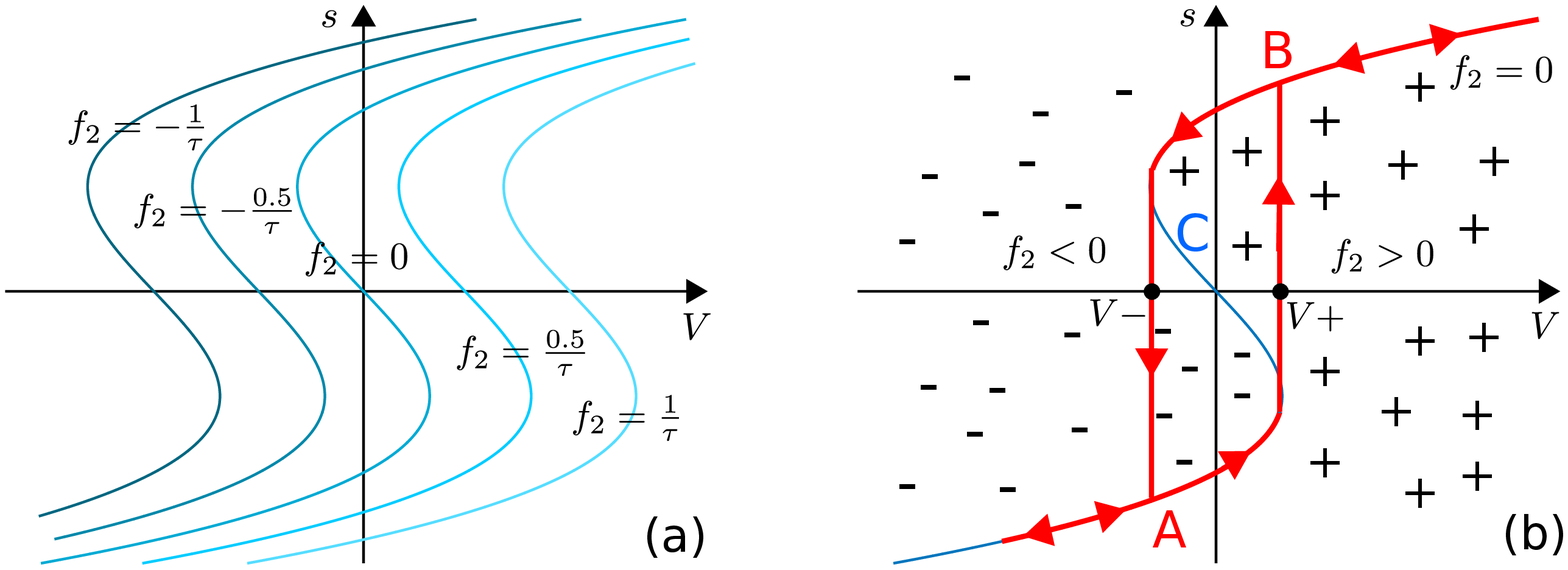,width=0.9\linewidth}
}
\caption{Contour plot of $f_2$ function in \eqnref{hys_f2} and predicted $s$--$v$
         hysteresis curve based on the sign of $f_2$.
    \figlabel{hys_f2}}
\end{figure}

Note that we are analyzing and predicting hysteresis based on the DC solution curve
defined by $f_2(v,~s) = 0$.
This clarifies a common confusion people have. 
As hysteresis is normally defined as a type of time-dependence between output
and input, people often believe that it has nothing to do with the circuit's or
device's DC properties.
It is true that hysteresis is normally observed in transient analysis.
But from the above discussions, we can see that it is indeed generated by the
multiple stability and the abrupt change in DC solutions.
As mentioned earlier, at a certain time $t$, $s(t)$ can be thought of as
encoding the memory of $v(t)$ from the past.
Its multiple stability reflects the different possible sets of history of
$v(t)$.
And the separation between $V+$ and $V-$ in the DC curves ensures that no
matter at what speed we sweep $v$, there will always be hysteresis in the
$s$--$v$ relationship.

When we sweep $v$ back and forth, curve \texttt{C}, the one with a negative
slope in \figref{hys_f2} (b) never shows up in solutions.
The reason is that, although it also consists of solutions of $f_2 = 0$, 
these solutions are not stable.
If a ($v$, $s$) point on curve \texttt{C} is perturbed to move above
\texttt{C}, whether because of physical noise or numerical error, it falls in
the $f_2 > 0$ region and will continue to grow until it reaches \texttt{B}.
Similarly, if it moves below \texttt{C}, it will decrease to curve \texttt{A}.
Therefore, it won't be observed during voltage sweep, leaving only \texttt{A}
and \texttt{B} to form the $s$--$v$ hysteresis curves.

\subsection{Compact Model in MAPP}\seclabel{hys_MAPP}

With the model equations for \texttt{hys\_example} defined in \eqnref{hys_f1}
and \eqnref{hys_f2}, how do we put them into a compact model so that we can
simulate it in circuits?
To answer this question, in this section, we first discuss our formulation of
the general form of device compact models, namely the ModSpec format
\cite{amsallem2011modspec,WaAaWuYaRoCICC2015MAPP}.
Then we develop the ModSpec model for \texttt{hys\_example} and implement it in MAPP
\cite{MAPPwebsite}.

ModSpec is MAPP's way of specifying device models.
A device model describes the relationship between variables using equations.
Among the variables of interest, some are the device's inputs/outputs; they are
related to the circuit connectivity.
We call them the device's I/Os.
In the context of electrical devices, they are branch voltages and currents.
Among all the I/Os, some may be expressed explicitly using the other variables;
they are the outputs of the model's explicit equations.
Furthermore, a device model can also have non-I/O internal unknowns and
implicit equations.
Taking all these possibilities into consideration, we specify model equations
in the following ModSpec format.
\begin{eqnarray}
   \vec z &=& \frac{d}{dt} \vec q_e(\vec x, \vec y) + \vec f_e(\vec x, \vec y, \vec u),
   \eqnlabel{ModSpec1}\\
   0 &=& \frac{d}{dt} \vec q_i(\vec x, \vec y) + \vec f_i(\vec x, \vec y, \vec u).
   \eqnlabel{ModSpec2}
\end{eqnarray}

Vectors $\vec x$ and $\vec z$ contain the device's I/Os: $\vec z$ comprises
those I/Os that can be expressed explicitly (for \texttt{hys\_example}, it
contains only $i$), while $\vec x$ comprises those that cannot (for
\texttt{hys\_example}, it is $v$).
$\vec y$ contains the model's internal unknowns (for \texttt{hys\_example}, it
is $s$), while $\vec u$ provides a mechanism for specifying time-varying inputs
within the device (\eg, as in independent voltage or current sources).
The functions $\vec q_e$, $\vec f_e$, $\vec q_i$ and $\vec f_i$ define the
differential and algebraic parts of the model's explicit and implicit
equations. 

For \texttt{hys\_example}, we can write its model equations in the ModSpec
format as follows.
\begin{equation}\eqnlabel{hys_ModSpec}
    \begin{split}
	&\vec f_e(\vec x, \vec y, \vec u) = \frac{\vec x}{R} \cdot (\tanh(\vec y)+1),
     \text{~~~} \vec q_e(\vec x, \vec y) = 0, \\
	&\vec f_i(\vec x, \vec y, \vec u) = 
	 \vec x - \vec y^3 + \vec y, \text{~~~} 
     \vec q_i(\vec x, \vec y) = - \tau \cdot \vec y,
    \end{split}
\end{equation}
with $\vec x = [v]$, $\vec y = [s]$, $\vec z = [i]$, $\vec u = []$.

We can enter the model information in \eqnref{hys_ModSpec} into MAPP by
constructing a ModSpec object \texttt{MOD}.
The code in \appref{hys_ModSpec_code} shows how to create this device model for
\texttt{hys\_example} entirely in the \MATLAB{} language.
For more detailed description of the ModSpec format, users can issue the
command ``\texttt{help ModSpec\_concepts}'' in MAPP.

\subsection{Simulation Results}\seclabel{hys_simulation}

In this section, we verify our analysis and prediction of $i$--$v$ hysteresis
in \secref{hys_eqn} by testing the compact models presented in
\secref{hys_MAPP} and in a circuit shown in \figref{vsrc_hys}.

\begin{figure}[htbp]
\centering{
    \epsfig{file=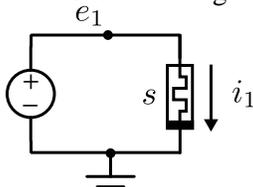,width=0.2\linewidth}
}
\caption{Schematic of the test bench circuit for \texttt{hys\_example}.
The three circuit unknowns are node voltage $e_1$, current $i_1$ and internal
state variable $s$.
    \figlabel{vsrc_hys}}
\end{figure}

\begin{figure}[htbp]
\centering{
    \epsfig{file=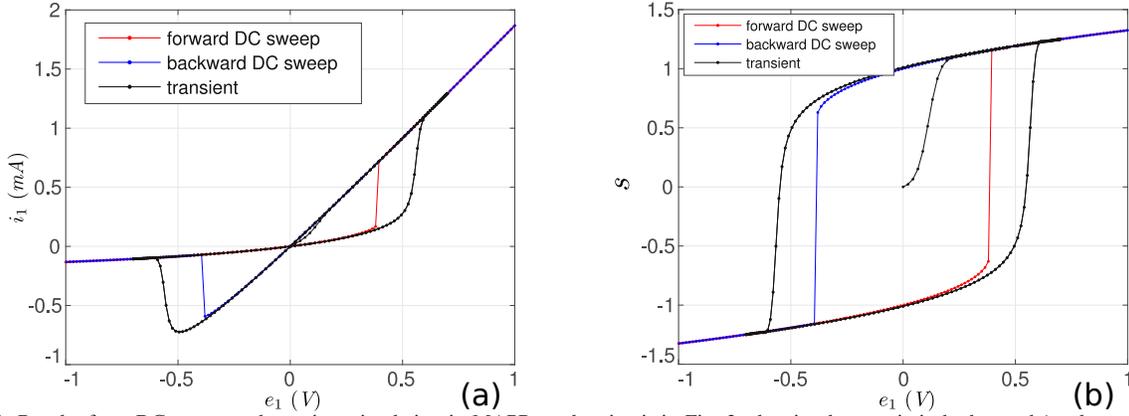,width=0.9\linewidth}
}
\caption{Results from DC sweep and transient simulation in MAPP on the circuit
in \figref{vsrc_hys}, showing hysteresis in both $s$ and $i_1$ when sweeping
the input voltage, in either type of the analyses.
    \figlabel{test_hys_sweep}}
\end{figure}

\figref{test_hys_sweep} shows the results from DC sweep and transient
simulation with input voltage sweeping up and down on the circuit in
\figref{vsrc_hys}.
It confirms that hysteresis takes place in both $i$--$v$ and $s$--$v$
relationships of the device.

\begin{figure}[htbp]
\centering{
    \epsfig{file=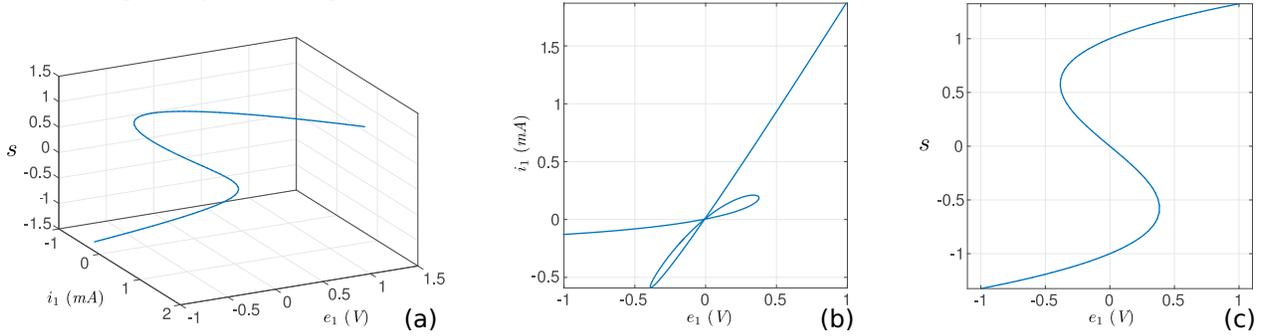,width=1.0\linewidth}
}
\caption{Results from homotopy analysis in MAPP.  (a) 3-D view of all the DC solutions of the circuit in
\figref{vsrc_hys} with voltage input between $-1${\it V} and $1${\it V}.
(b) top view of all the DC solutions shows the folding in the $i$--$v$
characteristic curve, explaining the $i$--$v$ hysteresis in \figref{test_hys_sweep}.
(c) side view of the DC solutions.
    \figlabel{test_hys_homotopy}}
\end{figure}

In \figref{test_hys_sweep} (b), curve \texttt{C} (defined in \figref{hys_f2}
(b)) with a negative slope never shows up in either forward or backward voltage
DC sweep.
This matches our discussion in \secref{hys_eqn}.
In order to plot this curve and complete the DC solutions, also to get
rid of the abrupt change of solutions in DC sweeps, we can use the homotopy
analysis \cite{RoMeTCAD2006}.
Homotopy analysis can track the DC solution curve in the state space.

Results from homotopy analysis on the circuit in \figref{vsrc_hys} are shown in
\figref{test_hys_homotopy}.
We note that all the circuit's DC solutions indeed form a smooth curve in the
state space.
The side view of the 3-D plot displays curve \texttt{C} we have designed in our
model equation \eqnref{hys_f2}.
The corresponding curve in the top view connects the two discontinuous DC sweep
curves in \figref{test_hys_sweep}; it consists of all the unstable solutions in
the $i$--$v$ relationship.
This curve was previously missing in DC and transient sweep results, and now
displayed by the homotopy analysis.
These results from homotopy analysis provide us with important insights into
the model.
They reveal that there is a single smooth and continuous DC solution curve in
the state space, which is an indicator of the well-posedness of the model.
They also illustrate that it is the folding in the smooth DC solution curve
that has created the discontinuities in DC sweep results.
These insights are important for the proper modelling of hysteresis. 

Moreover, the top view explains the use of internal state $s$ for modelling
hysteresis from another angle.
Without the internal state, it would be difficult if not impossible to write a
single equation describing the $i$--$v$ relationship shown in
\figref{test_hys_homotopy} (b).
With the help of $s$, we can easily choose two simple model equations as
\eqnref{hys_f1} and \eqnref{hys_f2}, and the complex $i$--$v$ relationship
forms naturally.

\section{\normalfont {\large How to Model Internal Unknowns Properly in Verilog-A}}
\seclabel{hys_VA}

In this section, we write the \texttt{hys\_example} model in the Verilog-A language.

Apart from the differences in syntax, Verilog-A differs from ModSpec in one
key aspect --- the way of handling internal unknowns and implicit equations.
Verilog-A models a device with an internal circuit topology, \ie, with internal
nodes and branches defined just like in a subcircuit.
The variables in a Verilog-A model, the ``sources'' and ``probes'', are
potentials and flows specified based on this topology.
Coming from this subcircuit perspective, the language doesn't provide a
straightforward way of dealing with general internal unknowns and implicit
equations inside the model, \eg, the state variable $s$ and the equation
\eqnref{sV_f2} in \texttt{hys\_example}.

This limitation gives rise to so much confusion about the modelling of devices
with hysteresis, that we would like to examine the common modelling mistakes and
pitfalls before describing our approach. 
Here is the list of how not to model internal unknowns and implicit equations
in Verilog-A.

\begin{list}{$\circ$ }
{
    \setlength{\topsep}{2pt}
    \setlength{\leftmargin}{10pt}
    \setlength{\labelsep}{1pt}
    \setlength{\rightmargin}{0pt}
    \setlength{\labelwidth}{10pt}
}
\item{
Declare the internal unknown as a general variable, \eg, using
``\texttt{real}'', then use ``\texttt{idt()}'' function to describe the
differential equation the variable should satisfy.
This approach is not recommended because of several reasons.

First, Verilog-A provides most consistent definitions and support for
potentials and flows as circuit unknowns; it is unclear how ``\texttt{real}''
variables inside differential equations are handled by each Verilog-A compiler.
Some simulators will return inconsistent or incorrect results.
Moreover, another potential hazard from this practice is that the simulator may
create a memory state for the variable \cite{WaRo2013VAnanoHUB}, limiting its
use in some simulation algorithms, \eg, those for periodic steady state (PSS)
analysis.

Also, people often attempt to use ``\texttt{idt()}'' in this scenario,
apparently because Verilog-A doesn't allow using ``\texttt{ddt()}'' to
contribute to a none-potential/flow quantity as ``source'', for good reasons.
But this ``workaround'' with the use of ``\texttt{idt()}'' is not recommended
\cite{WaRo2013VAnanoHUB,coram2004VA}, as different simulators have inconsistent
support for ``\texttt{idt()}''.
}
\item{
Another pitfall is to use implicit contributions.
While an implicit contribution in Verilog-A seems to simplify the code, and
forces users to model the internal unknown as a potential or flow, which is in
line with what we propose, it is not recommended
\cite{WaRo2013VAnanoHUB,mcandrew2015VA}.
In fact, it is not supported properly even by some well-known commercial
simulators.
}
\item{
Model the differential relationship by coding time integration inside.
In this approach, the model has access to the absolute time and calculates
the time step inside, then approximates the differential equation
\eqnref{sV_f2} by integrating $f_2$ at each time step.
The approach may seem straightforward, but it has so many problems that I have
to create another list for them:
\begin{list}{$\bullet$ }
{
    \setlength{\topsep}{2pt}
    \setlength{\leftmargin}{10pt}
    \setlength{\labelsep}{1pt}
    \setlength{\rightmargin}{0pt}
    \setlength{\labelwidth}{10pt}
}
\item{
The method inevitably uses ``\texttt{abstime}'' function in the model.
To set the starting point of the integration, it also has to use
the ``\texttt{initial\_step}'' event.
These are both bad practices in analog modelling
\cite{WaRo2013VAnanoHUB,coram2004VA}.
}
\item{
The method can only use Forward Euler (FE) \cite{RoNOWmonograph2009}
internally for integration, potentially causing convergence issues for stiff
systems.
}
\item{
In this method, the internal unknown is intentionally defined as a memory state,
again creating difficulties for PSS simulation.
}
\item{
The model won't perform correctly in analyses that do not involve time
integration, like DC, small signal AC analysis and Harmonic Balance.
}
\item{
Even for transient simulation, it 
defeats the purpose of using the simulator, as it bypasses the simulator's many
built-in facilities, \eg, convergence aiding techniques, truncation error
estimation, time step control, \etc.
}
\item{
There are many more issues with this approach.
For example, circuit designers cannot set transient analysis initial conditions for the
internal unknown the normal way they do for capacitor voltages and inductor
currents.
Also, to ``ensure'' the accuracy of internal time integration,
``\texttt{bound\_step}'' is often used.
And the bounded step specified either makes simulation inaccurate or
unnecessarily slow.
}
\end{list}
}
\end{list}

We note that these problems and pitfalls arise partly from the limitation of
the Verilog-A language in intuitively handling general internal unknowns and
implicit equations, mostly from bad modelling practices.
To circumvent these issues and write a robust Verilog-A model for
\texttt{hys\_example} that should work consistently in all simulators and all
simulation algorithms, we model state variable $s$ as a voltage.
We declare an internal branch, whose voltage represents $s$. 
One end of the branch is an internal node that doesn't connect to any other
branches.
In this way, by contributing $V - s^3 + s$ and \texttt{ddt(-tau * s)} both to
this same branch, the KCL at the internal node will enforce the implicit
differential equation in \eqnref{hys_f2}.

Declaring $s$ as a voltage is not the only way to model \texttt{hys\_example}
in Verilog-A.
Depending on the physical nature of $s$, one can also use Verilog-A's
multiphysics support and model it as a mechanical property, such as a position
from the kinematic discipline.
This may be closer to the actual meaning of $s$ for MIM-structured RAM devices.
Alternatively, we can also use the property for potential from the thermal or
magnetic discipline.
One can also switch potential and flow by defining $s$ as a flow instead.
These alternatives may make the model look more physical, but they do not make
a difference mathematically, except from the scale of tolerances in each
discipline, which we will discuss in more detail in \secref{RRAM_v0_VA}.
The essence of our approach is to recognize that state variable $s$ is a
circuit unknown, and thus should be modelled as a potential or flow in
Verilog-A, for the consistent support from different simulators in various
circuit analyses.

The Verilog-A code for \texttt{hys\_example} is provided in
\appref{hys_va_code}.
It generates consistent results in many simulation platforms, including 
\Spectre,\footnote{\Spectre{} version: 7.2.0 64bit.} 
\HSPICE,\footnote{\HSPICE{} version: J-2014.09 64bit.}
and the open-source simulator Xyce.\footnote{Xyce version: 6.4.}
The test benches with all these simulators can be found in \appref{hys}.

\section{\normalfont {\large RRAM Model}}\seclabel{RRAM}

The model \texttt{hys\_example} developed in \secref{hys} is a model template
for devices with hysteresis, such as RRAM devices.
By changing its $f_1$ \eqnref{hys_f1} and $f_2$ \eqnref{hys_f2} functions in
model equations, as well as the corresponding function implementations in MAPP
and Verilog-A code, we can then have compact models capturing the physics of
RRAM devices.

\subsection{Model Equations}\seclabel{RRAM_v0_eqn}

An RRAM device consists of two metal electrodes, namely $t$ (top) and $b$
(bottom), and a thin oxide film separating them.
A conductive filament can form in the film.
When it grows to connect the two electrodes, the device is in low resistance
state (LRS); when part of it dissolves, the device enters high resistance state
(HRS).
As a RAM, its ``memory'' is stored in the status of its internal conductive
filament and the corresponding resistance state.

From the above discussion, the internal state variable for RRAM models can be
either the length of the filament \cite{UMich2011RRAM}, 
or the gap between the tip of the filament and the opposing electrode
\cite{Stanford2012RRAMspice,Stanford2014RRAMverilog}.
We choose to use the gap in this section, as it is what really determines the
tunnelling current.
Then the variables in the RRAM model are: the voltage $vtb$ across the device,
the current $itb$ through it and the internal unknown $gap$.
We can then rewrite the equations \eqnref{IV_f1} and \eqnref{sV_f2} from the
model template in \secref{hys_eqn} as
\begin{eqnarray}
	itb(t) &=& f_1(vtb(t),~gap(t)), \eqnlabel{RRAM_f1} \\
	\frac{d}{dt} gap(t) &=& f_2(vtb(t),~gap(t)). \eqnlabel{RRAM_f2}
\end{eqnarray}

The physical contexts of these RRAM model equations are straightforward to
understand. 
Equation \eqnref{RRAM_f1} determines how the current is modulated by both the
voltage and $gap$;
equation \eqnref{RRAM_f2} describes the growth rate of $gap$ at a given voltage
with some existing $gap$ size.
Our goal of RRAM modelling is to find suitable $f_1$ and $f_2$ functions to capture
these physical properties.

The formula for $f_1$ are mostly consistent across several existing RRAM models
developed in different groups
\cite{UMich2011RRAM,Stanford2014RRAMverilog,ASU2015RRAMverilog,yakopcic2013memristor}.
Among them, 
\cite{Stanford2014RRAMverilog,ASU2015RRAMverilog} use the same equation, which
is only different from that used in \cite{yakopcic2013memristor} in the choice
of internal unknown.\footnote{In the I-V relationship equation in
\cite{yakopcic2013memristor}, we can redefine the internal unknown and make a
one-to-one mapping between $s^n$ and $\exp(-gap/g_0)$, to make the equation
equivalent to the one in \cite{Stanford2014RRAMverilog,ASU2015RRAMverilog}.}
Therefore, in this section, we choose to use the $f_1$ function in
\cite{Stanford2014RRAMverilog,ASU2015RRAMverilog}:
\begin{equation}\eqnlabel{RRAM_v0_f1}
	f_1(vtb,~gap) = I_0 \cdot \exp(-\frac{gap}{g_0}) \cdot \sinh(\frac{vtb}{V_0}),
\end{equation}
where $I_0$, $g_0$, $V_0$ are fitting parameters.

For $f_2$, we can adapt the $gap$ growth formulation in
\cite{Stanford2014RRAMverilog,ASU2015RRAMverilog} and write it as
\begin{equation}\eqnlabel{RRAM_v0_f2}
    f_2(vtb,~gap) = -v_0 \cdot \exp(- \frac{E_a}{V_T})\cdot
    \sinh(\frac{vtb \cdot \gamma \cdot a_0}{t_{ox} \cdot V_T}),
\end{equation}
where $v_0$, $E_a$, $a_0$ are fitting parameters, $t_{ox}$ is the thickness of
the oxide film, $V_T = k\cdot T/q$ is the thermal voltage, and
\begin{equation}\eqnlabel{RRAM_v0_gamma}
    \gamma = \gamma_0 - \beta \cdot gap^3.
\end{equation}

$\gamma$ in \eqnref{RRAM_v0_gamma} is known as the local field enhancement
factor \cite{mcpherson2003thermochemical}.
It accounts for the abrupt SET (filament grows enough to connect electrodes)
and gradual RESET (filament dissolves) behaviors in bipolar RRAM devices
\cite{yu2012neuromorphic}.
Parameters are normally chosen to ensure that this $\gamma$ factor is always
positive.
So the sign and zero-crossings of $f_2$ in \eqnref{RRAM_v0_f2} are determined
only by $vtb$.

While there are small differences among the $f_2$ functions in models developed
by various groups
\cite{UMich2011RRAM,Stanford2014RRAMverilog,ASU2015RRAMverilog,yakopcic2013memristor},
they differ mainly in the definitions of fitting parameters.
A property they all share is that the sign of $f_2$ is the same as that of
$-\sinh(vtb)$.
Put in other words, $gap$ begins to decrease whenever $vtb$ is positive, and
vice versa, as illustrated in \figref{RRAM_f2} (a).
While there is some physical truth to this statement, considering that an RRAM
device will eventually be destroyed\footnote{Under constant negative voltages,
the filament will dissolve to the extent that SET cannot restore it.
The device will need to go through the forming process again.
Under constant positive voltages, the filament will grow too thick to RESET,
and the device becomes shorted.} if applied a constant voltage for an
indefinite amount of time, for the model to work in numerical simulation, the
state variable $gap$ has to be bounded.

\begin{figure}[htbp]
\centering{
    \epsfig{file=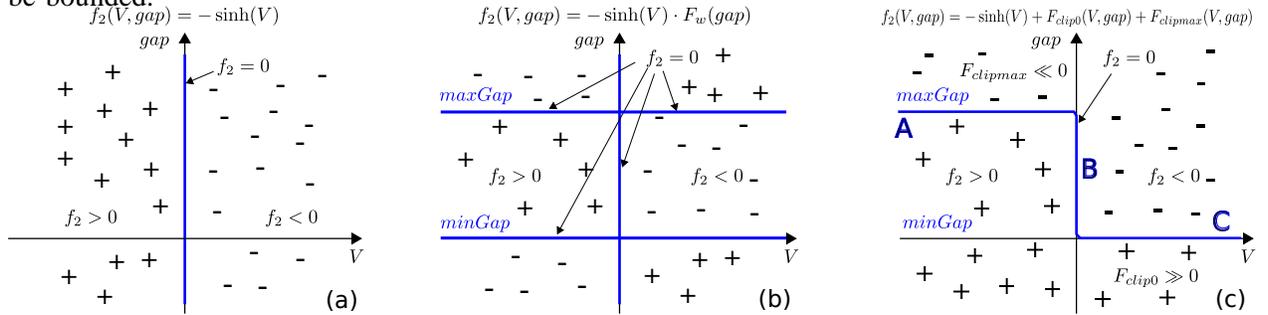,width=1.0\linewidth}
}
\caption{Illustration of several choices of $f_2$ in RRAM model.
    \figlabel{RRAM_f2}}
\end{figure}

Ensuring that the upper and lower bounds for $gap$ are always respected in
simulation is one major challenge for the compact modelling of RRAM devices.
To address this challenge, several techniques have been attempted in the
existing RRAM compact models:
\begin{list}{$\circ$ }
{
    \setlength{\topsep}{2pt}
    \setlength{\leftmargin}{10pt}
    \setlength{\labelsep}{1pt}
    \setlength{\rightmargin}{0pt}
    \setlength{\labelwidth}{10pt}
}
\item{
Directly use \texttt{if-then-else} statements on $gap$
\cite{Stanford2014RRAMverilog,ASU2015RRAMverilog}.
This type of model is normally written in Verilog-A.
They declare \texttt{gap} as a \texttt{real} variable, then directly enforce
``\texttt{if (gap < 0)  gap = 0;}''.
We have discussed in great detail in \secref{hys_VA} about the problems of
modelling internal unknowns as general Verilog-A variables.
On top of these problems, no matter whether the Verilog-A compiler treats
\texttt{gap} as a circuit unknown or a memory state, the use of
\texttt{if-then-else} statements for bounding the variable excludes the model
from the differential equation framework.
Thus they are not suitable for simulation analyses.

Moreover, the use of \texttt{if-then-else} also introduces hard discontinuities
in the model, causing convergence problems \cite{RoNOWmonograph2009}. 
Also, forcefully setting variable \texttt{gap} to certain values can result in
singular circuit Jacobian matrices, creating difficulties for most simulation
algorithms.
}
\item{
Use window functions
\cite{UMich2011RRAM,yakopcic2013memristor}.

The goal is to set $\frac{d}{dt} gap = f_2 = 0$ when $gap = maxGap$ and $gap =
minGap$.
The method used in these models is to multiply the $f_2$ in \eqnref{RRAM_v0_f2}
with a window function that is close to $1$ when $minGap < gap < maxGap$, equal
to $0$ when $gap$ is at $minGap$ or $maxGap$, and has negative values
elsewhere.
Directly constructing such windows functions with \texttt{step()} functions
\cite{UMich2011RRAM} is not recommended as it introduces discontinuities into
the model.
One better example of such window functions for $[0, 1]$ window size is known
as the Joglekar window \cite{yakopcic2014memristor}:
\begin{equation}\eqnlabel{RRAM_v0_Fw}
	F_w(x) = 1- (2\cdot x-1)^{2\cdot p},
\end{equation}
where $p$ is a positive integer used to adjust the sharpness of the window.

After multiplying window functions, the $f_2$ function used in these models is
still smooth and continuous, and the models still in the differential equation
format, complying with the model template we have discussed in \secref{hys}.
As a result, the models are often reported to run reasonably well in transient
simulations
\cite{kvatinsky2013team,yakopcic2014memristor,kvatinsky2011memristor}.

However, there are subtle and deeper problems with this approach. 
The problems can also be illustrated by analyzing the sign and zero-crossings
of function $f_2$.
After multiplying $f_2$ by window functions, the zero-crossings of $f_2$ are
shown in \figref{RRAM_f2} (b).
The $f_2 = 0$ curves consist of three lines: the $maxGap$ and $minGap$ lines,
and the $V=0$ line.
Based on the sign of $f_2$, the left half of the $minGap$ line and the right
half of the $maxGap$ line consist of unstable DC solutions; they are unlikely
to show up in transient simulations.
Therefore, when sweeping the voltage between negative and positive values,
$gap$ will move between $maxGap$ and $minGap$.
This is the foundation for the model to work in transient simulations.
However, based on \figref{RRAM_f2} (b), the model has several problems 
in other types of analyses.
\begin{list}{$\bullet$ }
{
    \setlength{\topsep}{2pt}
    \setlength{\leftmargin}{10pt}
    \setlength{\labelsep}{1pt}
    \setlength{\rightmargin}{0pt}
    \setlength{\labelwidth}{10pt}
}
\item{
In DC operating point analysis or DC sweeps, all lines consisting the $f_2=0$
curves can show up, including those containing unphysical results.
For example, when the voltage is zero, any $gap$ size is a solution; $gap$ is
not bounded anymore. 
}
\item{
In homotopy analysis, the intersection of solution lines introduced by the
window functions makes the solution curve difficult to track.
In particular, it will attempt to track the $V=0$ line where $gap$ grows
without bound.
The fact that there is no single continuous solution curve in the state space
indicates poor numerical properties of the model in other types of simulation
algorithms as well.
}
\item{
Even in transient analysis, the model won't run properly unless we carefully
set an initial condition for $gap$.
If the initial value of $gap$ is beyond $(minGap,~maxGap)$, or if it falls
outside this range due to any numerical error, it can start to grow without
bound.
}
\end{list}
Other window functions are also tried for this approach, \eg, Biolek and
Prodromakis windows
\cite{yakopcic2014memristor,kvatinsky2011memristor}.
But as long as the window function is multiplied to $f_2$, the picture of DC
solutions in \figref{RRAM_f2} (b) stays the same.
And it is this introduction of unnecessary DC solutions the modelling artifact
that limits the RRAM model's use in simulation analyses.
}
\end{list}

In our approach, we try to bound variable $gap$ while keeping the DC solutions
in a single continuous curve, illustrated as the $f_2 = 0$ curve in
\figref{RRAM_f2} (c).
This is inspired by studying the model template \texttt{hys\_example} in
\secref{hys}.
The sign and zero-crossing of $f_2$ for our RRAM model are closely related to
those of the $f_2$ function \eqnref{hys_f2} for \texttt{hys\_example} (shown in
\figref{hys_f2}).

The desired $f_2=0$ solution curve consists of three parts: curve \texttt{A}
and \texttt{C} contain the stable solutions; curve \texttt{B} contains those
that are unstable (or marginally stable).
In this way, when sweeping the voltage past zero, variable $gap$ will start to
switch between $maxGap$ and $minGap$.
If the sweeping is fast enough, I-V hysteresis will show up.

To construct the desired $f_2=0$ solution curve, we modify the original $f_2$ in
\eqnref{RRAM_v0_f2} by adding clipping terms to it.
Our new $f_2^*$ can be written as
\begin{equation}\eqnlabel{RRAM_v0_f2star}
    f_2^*(vtb,~gap) = f_2(vtb,~gap) + F_{clipmin}(vtb,~gap) + F_{clipmax}(vtb,~gap), 
\end{equation}
where $f_2$ is the original function in \eqnref{RRAM_v0_f2}, $F_{clipmin}$ and
$F_{clipmax}$ are clipping functions:
\begin{eqnarray}
	F_{clipmin}(vtb,~gap) &=& (\text{safeexp}(K_{clip} \cdot (minGap-gap),~maxslope)
    - f_2(vtb,~gap))
    \cdot F_{w1}(gap),\eqnlabel{RRAM_v0_Fmin} \\
	F_{clipmax}(vtb,~gap) &=& (-\text{safeexp}(K_{clip} \cdot (gap-maxGap),~maxslope)
    - f_2(vtb,~gap)
    \cdot F_{w2}(gap).\eqnlabel{RRAM_v0_Fmax}
\end{eqnarray}

Functions $F_{w1}$ and $F_{w2}$ in \eqnref{RRAM_v0_Fmin} and \eqnref{RRAM_v0_Fmax} are
smooth versions of step functions:
\begin{eqnarray}
    F_{w1}(gap) &=& \text{smoothstep}(minGap-gap, ~~ smoothing), \eqnlabel{RRAM_v0_Fw1} \\
    F_{w2}(gap) &=& \text{smoothstep}(gap-maxGap, ~~ smoothing). \eqnlabel{RRAM_v0_Fw2}
\end{eqnarray}

The intuition behind $F_{w1}$ and $F_{w2}$ is to make $F_{w1} \approx 0$ and 
$F_{w2} \approx 0$ when $gap$ is within $[minGap,~maxGap]$;
then $F_{w1} \approx 1$ when $gap < minGap$, $F_{w2} \approx 1$ when $gap >
maxGap$.

When $F_{w1} \approx 1$ or $F_{w2} \approx 1$, the added clipping term in
\eqnref{RRAM_v0_Fmin} or \eqnref{RRAM_v0_Fmax} is ``in effect''.
Either term will first use $-f_2(vtb,~gap)$ to cancel out the effect of $f_2$,
then add a fast growing component modelled using exponential functions to ensure
that $f_2^*$ has the desired sign as in \figref{RRAM_f2} (c).
Parameter $K_{clip}$ is used to adjust the speed in which these exponential
components grow.

Note that in equations \eqnref{RRAM_v0_Fmin}, \eqnref{RRAM_v0_Fmax} and
\eqnref{RRAM_v0_Fw1}, \eqnref{RRAM_v0_Fw2}, instead of using normal exponential
and step functions, we use $\text{safeexp()}$ and $\text{smoothstep()}$.
These are smooth functions we have developed with better numerical properties
than the original ones.
$\text{safeexp()}$ linearises the exponential function from the point its
derivative reaches parameter $maxslope$.
$\text{smoothstep()}$ is implemented whether as a parameterised $\tanh$, or as
\begin{equation}\eqnlabel{smoothstep}
    \text{smoothstep}(x) = 0.5 \cdot (\frac{x}{\sqrt{x^2 + smoothing}}+1).
\end{equation}

Issuing commands ``\texttt{help safeexp;}'' and ``\texttt{help smoothstep;}''
in MAPP will display more usage and implementation details of these functions.

The $f_2^*$ we have proposed for RRAM model is smooth and continuous in both
$vtb$ and $gap$.
Its sign and zero-crossings are designed to mimic those shown in
\figref{RRAM_f2} (c).
By adjusting the parameters $K_{clip}$ and $smoothing$, users can tune the
sharpness of the DC solution curve in \figref{RRAM_f2} (c).
The clipping terms can also leave the values from the original $f_2$ function
in \eqnref{RRAM_v0_f2} almost intact when $minGap < gap < maxGap$.

While the intention of adding the clipping terms in \eqnref{RRAM_v0_f2star}
is to set up bounds for variable $gap$ and to construct DC solution curve in
\figref{RRAM_f2} (c), there is also some physical justification to our
approach.
As a physical quantity, $gap$ is indeed bounded by definition.
Therefore, $\frac{d}{dt} gap = f_2$ cannot look like \figref{RRAM_f2} (a) in
reality.
The $f_2 = 0$ curves must have the \texttt{A} and \texttt{B} parts in
\figref{RRAM_f2} (c).
One can think of the clipping terms as infinite amount of resisting ``force''
to keep $gap$ from decreasing below $minGap$, or increasing beyond $maxGap$.
The analogy is the modelling of MEMS switches, where the switching beam's
position is often used as an internal state variable.
This variable reaches its bound when the switching beam hits the opposing
electrode (often the substrate).
The position does not move further.
The beam cannot move into the electrode/substrate because of the huge force
resisting it from causing any shape change in the structures. 
Similarly, in RRAM modelling, if the variable $gap$ represents it physical
meaning accurately, one can expect such ``forces'' to exist to make
it a bounded quantity.
This physics intuition matches well with our proposed numerical technique of
using fast growing exponential components to enforce the bounds.

The compact model we propose for RRAM devices, with equations
\eqnref{RRAM_v0_f1} and \eqnref{RRAM_v0_f2star}, complies with the differential
equation format.
It uses the correct model template for hysteretic devices proven to work.
The study of the model template and the use of it for RRAM help us avoid many
of the modelling pitfalls at this equation formulation stage.
Compared with existing models, our model does not have to use
``\texttt{idt()}'' \cite{Stanford2014RRAMverilog,li2015RRAM}, or
events and functions like ``\texttt{initial\_step}'', ``\texttt{bound\_step}''
and ``\texttt{abstime}'' \cite{Stanford2014RRAMverilog,ASU2015RRAMverilog}.
It is not limited to using SPICE subcircuits written in simulator-dependent
syntax \cite{UMich2011RRAM,yakopcic2013memristor}.
With our model formulation, for the first time, it is possible to write robust
compact models for RRAM devices in both ModSpec and Verilog-A, that should run
consistently on various simulation platforms in different analyses.

Apart from the use in modelling and simulation, our analysis of the RRAM
equations provides important insights into the physical nature of these
devices.
Comparing \figref{hys_f2} (b) and \figref{RRAM_f2} (c), we note that the $f_2$
function for RRAM, unlike that of the model template \texttt{hys\_example},
does not have DC solutions folding back with a negative slope.
We can say that there is no ``DC hysteresis'' for these devices.
Put in other words, if voltage is swept slowly enough, there will be no I-V
hysteresis;
there will only be an abrupt change in $gap$ at zero voltage.
We would like to clarify that this does not constitute a problem for using
RRAMs as memory devices.
Because the growth rate of filament is exponential in the input voltage; only
when the voltage is substantially large will the growth be significant.
When the applied voltage is small, it may take years or decades for SET and
RESET to happen. Therefore, the device can still keep its ``memory'' securely.
From our analysis, it is this exponential relationship that accounts for the
switching voltages measured in RRAM devices.
But the lack of ``DC hysteresis'' distinguishes RRAM from the general
hysteresis devices like \texttt{hys\_example}.
This provides new perspective to the debate over whether RRAMs are
memristors or not
\cite{meuffels2012RRAMnotMemristor,di2013RRAMnotMemristor,slipko2013RRAMnotMemristor}.
The lack of ``DC hysteresis'' in RRAM devices explains why they cannot cope
with inevitable thermal fluctuations and will erratically change state over
time in the presence of noise \cite{slipko2013RRAMnotMemristor}.
Although showing I-V hysteresis curves like a genuine memristor during voltage
sweeps, RRAMs are more like ``chemical capacitors'' as they violate some
essential requirements on a genuine memristor \cite{di2013RRAMnotMemristor}.
It is arguable whether these criticisms are valid.
Nevertheless, our analysis in this section explains the difference between
\texttt{hys\_example}, a device with true ``DC hysteresis'' and the RRAM device
model vigorously, while being easy to appreciate graphically. 

\subsection{Compact Model in MAPP}\seclabel{RRAM_v0_MAPP}

Similar to the hys model in \secref{hys}, we can put the RRAM equations
$f_1$ \eqnref{RRAM_v0_f1} and $f_2^*$ \eqnref{RRAM_v0_f2star} into a compact
model by writing them in the ModSpec format:
\begin{equation}\eqnlabel{RRAM_v0_ModSpec}
    \begin{split}
	&\vec f_e(\vec x, \vec y, \vec u) = f_1(\vec x,~\vec y),
     \text{~~~} \vec q_e(\vec x, \vec y) = 0, \\
	&\vec f_i(\vec x, \vec y, \vec u) = 
	 f_2^*(\vec x,~\vec y), \text{~~~} 
     \vec q_i(\vec x, \vec y) = - 10^{-9} \cdot \vec y,
    \end{split}
\end{equation}
with $\vec x = [vtb]$, $\vec y = [gap]$, $\vec z = [itb]$, $\vec u = []$.

Note that there is $10^{-9}$ in the $\vec q_i$ function.
This is to scale the equation for better convergence.
We explain this technique in more detail in \secref{RRAM_v0_VA}.

The code in \appref{RRAM_v0_ModSpec_code} shows how to enter this RRAM model
into MAPP.

\subsection{Compact Model in Verilog-A}\seclabel{RRAM_v0_VA}

Having followed the model template discussed in \secref{hys} and formulated the
RRAM model in the differential equation format in \secref{RRAM_v0_eqn}, in this
section, we discuss the Verilog-A model for RRAM.
The Verilog-A model is show in \appref{RRAM_v0_va_code}.

Same as in the Verilog-A model for \texttt{hys\_example} (\secref{hys_VA}), we
also model the internal state variable $gap$ in RRAM as a voltage.
We have discussed why this approach results in more robust Verilog-A models
compared with many alternatives, \eg, using ``\texttt{idt()}''
\cite{Stanford2014RRAMverilog,li2015RRAM}, implementing time integration inside
models \cite{ASU2015RRAMverilog}, \etc.
In this section, we would like to highlight from the provided Verilog-A code a
few more details in our modelling practices.
\begin{list}{$\circ$ }
{
    \setlength{\topsep}{2pt}
    \setlength{\leftmargin}{10pt}
    \setlength{\labelsep}{1pt}
    \setlength{\rightmargin}{0pt}
    \setlength{\labelwidth}{10pt}
}
\item{
\uline{\it{Scaling of unknowns and equations.}}
In the Verilog-A code, we can see that $gap$ is modelled in nano-meters, as
opposed to meters.
This is not an arbitrary choice; the intention is to bring the value of this
variable to around $1$, at the same scale as other voltages in the circuit.
When the simulator solves for an unknown, only a certain accuracy can be
achieved, controlled by absolute and relative tolerances.
The abstol in most simulator for voltages is set to be $10^{-6}$V.
If gap is modelled in meters with nominal values around $10^{-9}$, it won't be
solved accurately.
Apart from the scaling of unknowns, we can also see from the Verilog-A code
another $10^{-9}$ factor in the implicit equation, scaling down its value.
In this RRAM model, the implicit equation is represented as the KCL at the
internal node.
The equality in KCL is calculated to a certain accuracy as well --- often
$10^{-12}$A.
However, without scaling down, the equation is expressed in nano-meter per
second.
For RRAM models, this is a value around $10^6$.
The simulator has to ensure an accuracy of at least 18 digits such that the KCL
is satisfied, which is not necessary and often not achievable with double
precision.
So we scale it by $10^{-9}$ to bring its nominal value to around $10^{-3}$,
just like a regular current in a circuit. 

Note that when explaining the scaling of unknowns and equations, we are using
the units nm or nm/s, mainly for readers to grasp the idea more easily.
It doesn't indicate that certain units are more suitable for modelling than
others.
The essence of scaling is to make the model work better with simulation
tolerances set for unknowns and equations.
}
\item{
\uline{\it{Numerical accuracy.}}
Note that in the Verilog-A code, we include the standard
\texttt{constants.vams} file and use physical constants from it.
This practice ensures that we are using these constants with their best
accuracy; their values will also be consistent with other models also including
\texttt{constants.vams}.
Although this is straightforward to understand, it is often neglected in
existing models.
For example, in the model released in \cite{UMich2011RRAM}, many constants are
used with only two digits of accuracy.
A variable named alpha, which can be calculated with 16 digits, is hard-coded
to $1.4\times 10^{19}$.
Since numerical errors propagate through computations, the best accuracy the
model can possibly achieve is limited to two digits, and worse if the
inaccurate variables are used in non-linear functions.
}
\item{
\uline{\it{Smooth and safe functions.}}
In the Verilog-A code, we have used \texttt{limexp}, \texttt{smoothstep}.
As discussed earlier, these functions help with convergence greatly and are
highly recommended for use in compact models.
}
\end{list}

\subsection{Simulation Results}\seclabel{RRAM_v0_simulation}
In this section, we simulate the RRAM model in a test circuit with the same
schematic as in \figref{vsrc_hys}.
The transient simulation results are shown in \figref{RRAM_v0_tran}, with the
I-V relationship plotted in log scale in \figref{RRAM_v0_tran} (b).
The results clearly show pinched hysteresis curves.

\begin{figure}[htbp]
\centering{
    \epsfig{file=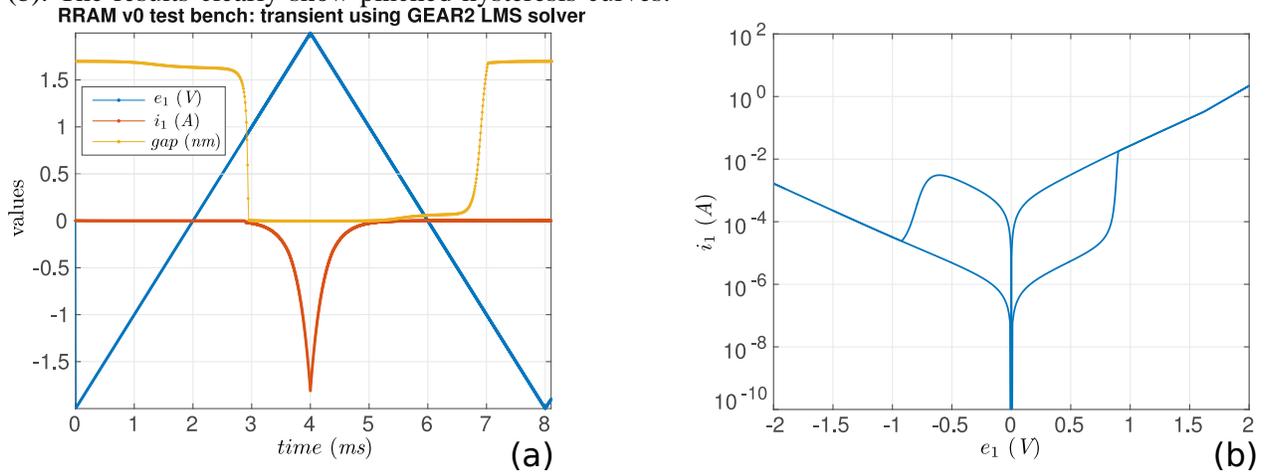,width=1.0\linewidth}
}
\caption{Transient results on the circuit with a voltage source connected to an
RRAM device.
    \figlabel{RRAM_v0_tran}}
\end{figure}

\begin{figure}[htbp]
\centering{
    \epsfig{file=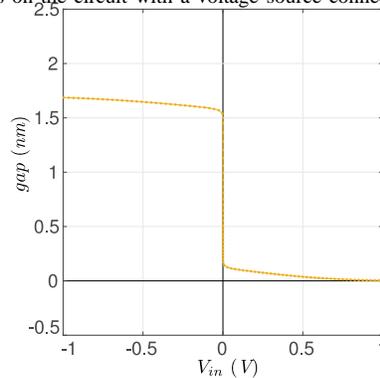,width=0.3\linewidth}
}
\caption{Homotopy analysis results on the circuit with a voltage source
connected to an RRAM device.
    \figlabel{RRAM_v0_homotopy}}
\end{figure}

The model we develop also work in DC and homotopy analyses. $gap$-$V$
relationship under DC conditions acquired from homotopy analysis are shown in
\figref{RRAM_v0_homotopy}.
DC sweeps from both directions in this case give the same results since the
model doesn't have DC hysteresis.
The $gap$-$V$ curve in \figref{RRAM_v0_homotopy} matches our discussion on the
$f_2 = 0$ solutions in \secref{RRAM_v0_eqn}.

Note that in the transient results, gap is not perfectly flat at $minGap$ or
$maxGap$; same phenomenon can also be observed in the DC solutions obtained
using homotopy.
This is because that the clipping functions we use, although fast growing,
cannot set exact hard limits on the internal unknown.
In other words, even when gap is close to $minGap$ or $maxGap$, changing the
voltage can still affect gap slightly.
This is not a modelling artifact.
In fact, this makes the model numerically robust, and at the same time more
physical.
It maintains the smoothness of equations and reduces the chance for Jacobian
matrix to become singular in simulation.
Physically, even when $gap$ is close to the boundary, changing voltage still
causes the device's state to change.
The small changes in $gap$ in this scenario can be interpreted as reflecting
the change in device's state, \eg, the width of the filament.
We conclude that, by making the model equations smooth, we are actually making
the model more physical.

\section{\normalfont {\large Convergence Aids}}\seclabel{convergence}

A common issue with newly-developed compact models of non-linear devices is
that they often do not converge in simulation.
In this section, we discuss several techniques in compact modelling that can
often improve the convergence of simulation.
Among these techniques, we focus on the use of SPICE-compatible limiting
functions.
We explain the intuition behind this technique and use this intuition to design
a limiting function specific to the RRAM model.

In the previous sections, we have already discussed several convergence aiding
techniques used in our RRAM model.
One of them is the proper scaling of both unknowns and equations.
This improves both the accuracy of solutions and the convergence of simulation.
The use of GMIN makes sure that the two terminals are always connected with a
finite resistance, reducing the chance for the circuit Jacobian matrix to
become singular during simulation.
We have also discussed the use of smooth and safe functions
(\texttt{smoothstep()}, \texttt{safeexp()}).
We highly recommend that compact model developers consider these techniques
when they encounter convergence issues with their models.

However, the above techniques do not solve all the convergence problems with
the RRAM model.
In particular, we have observed that the values and derivatives of
$f_1$ \eqnref{RRAM_v0_f1} and $f_2$ \eqnref{RRAM_v0_f2} often become very large
while the Newton Raphson (NR) iterations \cite{RoNOWmonograph2009} are trying
different guesses during DC operating point analysis.
This is because of the fast-growing sinh functions in the equations.
One solution is to use safesinh instead of sinh.
The safesinh function uses safeexp/limexp inside to eliminate the fast-growing
part with its linearized version, keeping the function values from exploding
numerically.
Although it has some physical justifications, it also has the potential
problems of inaccuracy, especially since the exponential relationship is the
key to the switching behaviour of RRAM devices (\secref{RRAM_v0_eqn}).
Therefore, in this section, we focus on another technique that can keep the
fast-growing exp or sinh function intact, but prevent NR from evaluating these
functions with large input values.
The techniques are known as initialization and limiting; they were implemented
in Berkeley SPICE, for nonlinear devices such as diodes, BJTs and MOSFETs.
Initialization evaluates these fast-growing nonlinear equations of
semiconductor devices with ``good'' voltage values at the first NR iteration;
limiting changes the NR guesses for these voltages in the subsequent
iterations, based on both the current guess at each iteration and the value
used in the last evaluation.

The limiting functions in SPICE include \texttt{pnjlim}, \texttt{fetlim} and
\texttt{limvds}.
Among them, \texttt{pnjlim} calculates new p-n junction voltage based on the
current NR guess and the last junction voltage being used, in an attempt to
avoid evaluating the exp function in the diode equation with large values.
This mechanism is applicable to sinh as well.
Inspired by \texttt{pnjlim}, we design a \texttt{sinhlim} that can reduce the
chance of numerical exposion for the RRAM model.

\begin{figure}[htbp]
    \begin{minipage}{0.44\linewidth}
      \epsfig{file=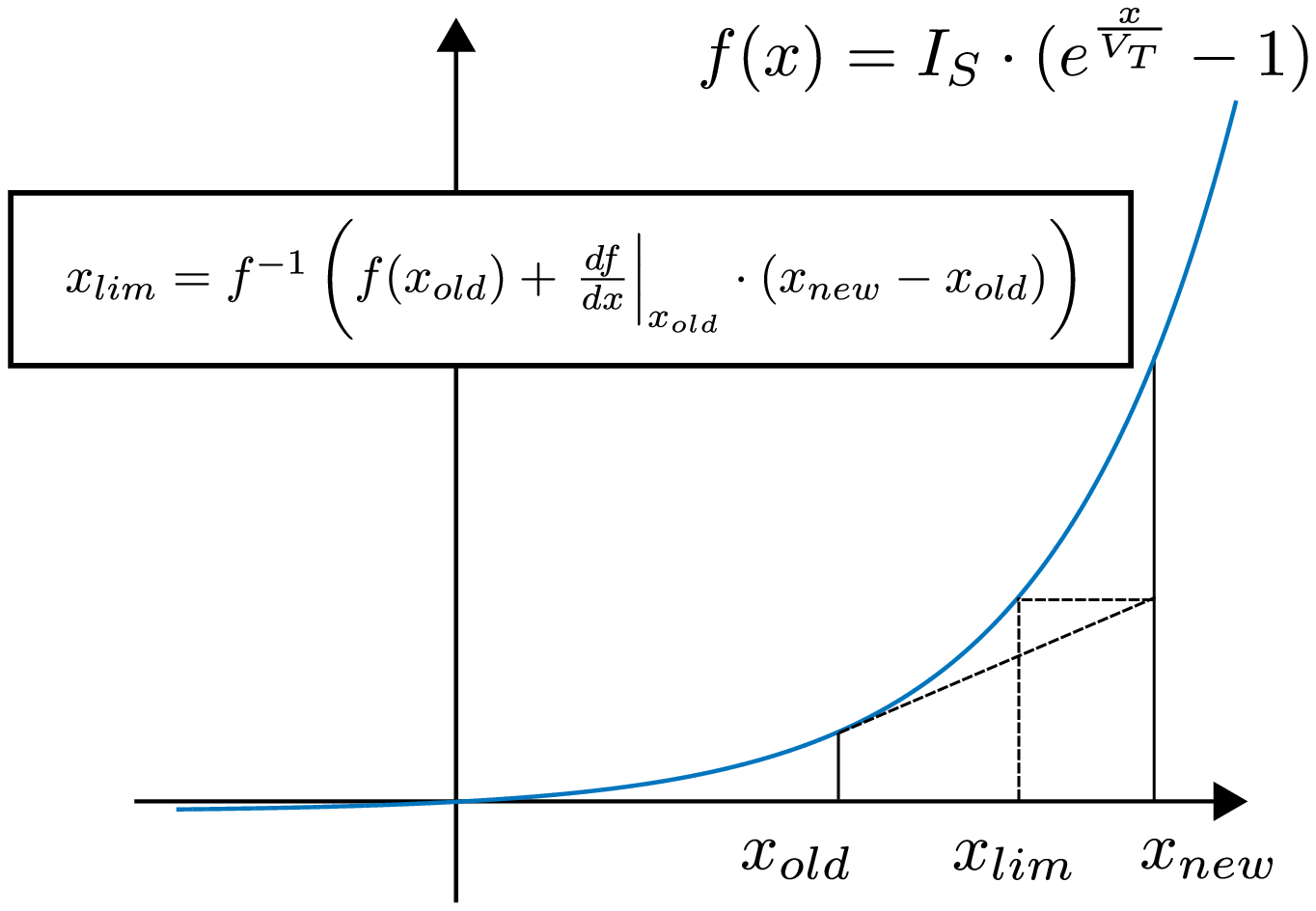,width=\linewidth}
      \caption{Illustration of pnjlim function in SPICE3.
       }\figlabel{pnjlim}
    \end{minipage}
    \hfill
    \begin{minipage}{0.41\linewidth}
      \epsfig{file=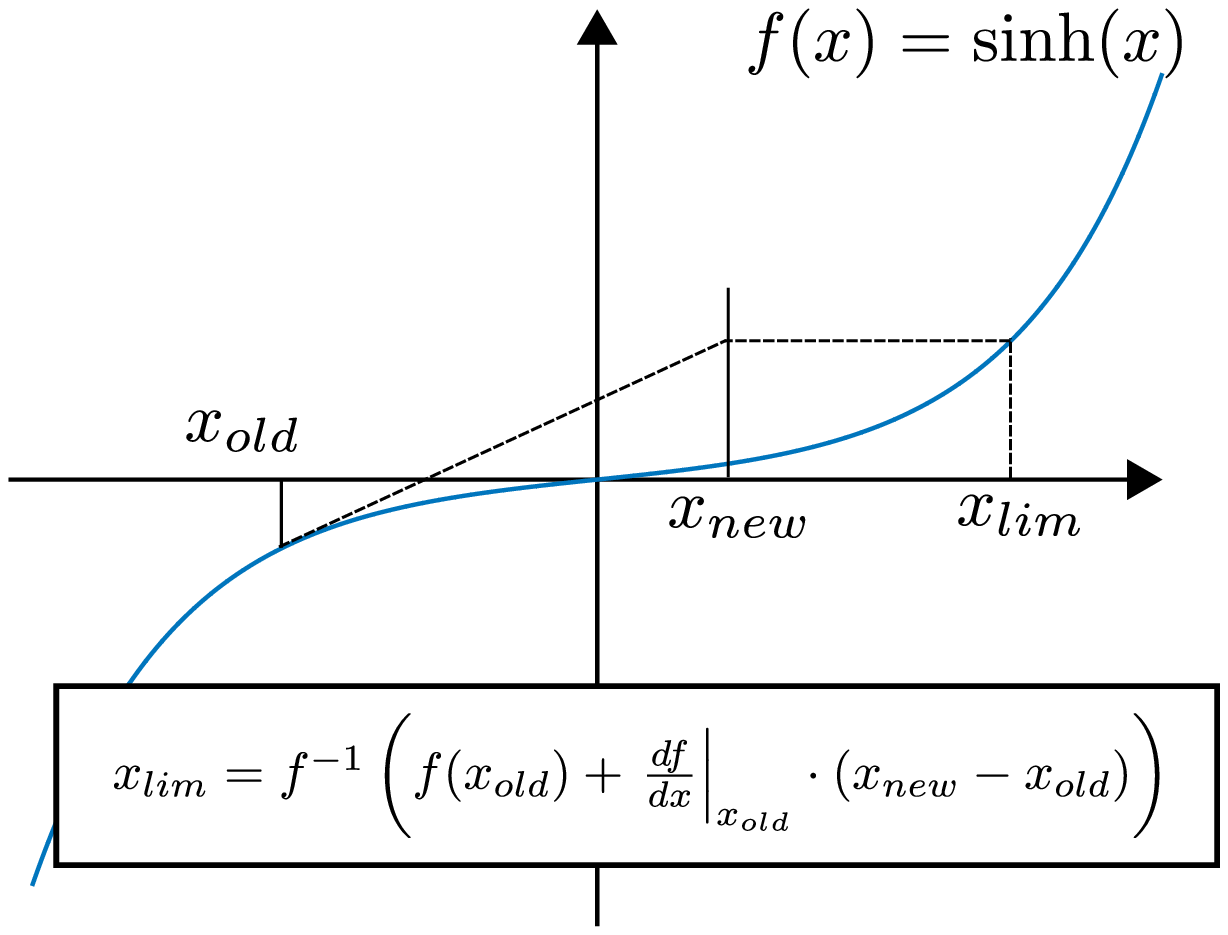,width=\linewidth}
	  \caption{Illustration of sinhlim function in MAPP.
	   }\figlabel{sinhlim}
    \end{minipage}
\end{figure}

\texttt{pnjlim} calculates the new junction voltage using the mechanism
illustrated in \figref{pnjlim}.
The current NR guess is $x_{new}$,  which is too large a value for evaluating
an exponential function.
So \texttt{pnjlim} calculates the limited version, $x_{lim}$, in between
$x_{new}$ and $x_{old}$.
Since NR linearized the system equation at $x_{old}$ in the last NR
iteration, and the linearization indicates that the new guess is $x_{new}$,
what NR actually wants is for the p-n junction to generate the current predicted
for $x_{new}$.
Because this prediction is based on the linearization at $x_{old}$, the actual
current at $x_{new}$ is apparently far larger than it.
Therefore, a more sensible choice for the junction voltage should be one that
gives out the predicted current.
From the above discussion, we can write an equation for the desired $x_{lim}$:
\begin{equation}
    I_S \cdot (e^\frac{x_{lim}}{V_T} - 1) =
	y_{lim} = I_S \cdot (e^\frac{x_{old}}{V_T} - 1) +
    I_S \cdot \frac{1}{V_T} \cdot e^\frac{x_{old}}{V_T}
    \cdot (x_{new} - x_{old}).
\end{equation}

Solving $x_{lim}$ from the above equation, we get the core of \texttt{pnjlim}.
\begin{equation}
    x_{lim} = \text{pnjlim\_core}(x_{new},~x_{old},~V_T) = 
    x_{old} + V_T \cdot \ln \left( 1 + \frac{x_{new} - x_{old}}{V_T} \right).
\end{equation}

From the above formula, the operation of \texttt{pnjlim} is essentially
inverting the diode I-V equation to calculate the desired voltage from the
predicted current at $x_{old}$.
Based on the same idea, we can write the limiting function for sinh.
As illustrated in \figref{sinhlim}, given $x_{old}$ and the current guess
$x_{new}$, we can calculate the desired ``current'' (function value), then
invert sinh to get the corresponding $x_{lim}$ for function evaluation.
Such an $x_{lim}$ satisfies
\begin{equation}
    \sinh(x_{lim}) = y_{lim} = \sinh(k \cdot x_{old}) + k \cdot \cosh(k \cdot x_{old})
    \cdot (x_{new} - x_{old}),
\end{equation}
which gives out the formulation of \texttt{sinhlim}: 
\begin{equation}
    x_{lim} = \text{sinhlim}(x_{new},~x_{old},~k) = \frac{1}{k} \cdot \ln\left(y_{lim} + \sqrt{1+y_{lim}^2}\right).
\end{equation}

This new limiting function \texttt{sinhlim} can be easily implemented in any
SPICE-compatible circuit simulator.
To demonstrate its effectiveness, we implement a simple two-terminal device
with its I-V relationship governed by a sinh function, \ie, the device equation
is $I = \sinh(V)$.
As sinh is a rapidly-growing function, even a simple circuit with a series
connection of a voltage source, a resistor of $1\Omega$ and this device may not
converge if the supply voltage is large.
This is because when searching for the solution, plain NR algorithm may try
large voltage values as inputs to the model's sinh function, resulting
difficulties or failure in convergence.
In contrast, SPICE-compatible NR can use \texttt{sinhlim} to calculate
$x_{lim}$ for use in iterations, preventing using large $x_{new}$ directly.
We run DC operating point analyses on this simple circuit, with NR starting
from all-zeros as initial guesses.
As shown in \tabref{sinhIV}, with the same convergence criteria,\footnote{In
the simulation experiments, reltol is 1e-6, abstol is 1e-12, residualtol is
1e-12.} the use of \texttt{sinhlim} improves convergence greatly.

\begin{table}[htb]
    \begin{center}
        \begin{tabular}{|c|c|c|}
        \hline
        {\bf Supply Voltage (V)} &
        {\bf with \texttt{sinlim} (niters)} &
        {\bf without limiting (niters)} \\ \hline
        \hline
        1 & 4 & 4
        \\ \hline
        10 & 4 & 9
        \\ \hline
        100 & 4 & 50
        \\ \hline
        1000 & 4 & \scriptsize{non-convergence within 100 iters}
        \\ \hline
        \end{tabular}
    \end{center}
  \caption{\normalfont{Number of NR iterations required for DC operating point
  analyses on a circuit with a series connection of a voltage source, a
resistor and a device with sinh I-V relationship.} \tablabel{sinhIV}}
\end{table}

We implement parameterized versions of the \texttt{sinhlim} function in our
RRAM model to aid convergence; the code is included in
\appref{RRAM_v0_ModSpec_code}.
Since there are two sinh function used in the RRAM model, in both $f_1$ and
$f_2$, two limited variables are declared in the model, with two
\texttt{sinhlim} with different parameters used in a vectorized limiting
function.

Many simulators available today are SPICE-compatible, in the sense that they
implement the equivalent limiting technique as in SPICE.\footnote{Some
simulators implement a technique known as ``limiting correction'', which is
compatible with SPICE's limiting formulation. Our \texttt{sinhlim} can also be
incorporated there.}
However, we would like to note that the limiting functions available in
literature today, 40 years after the introduction of SPICE, are still limited
to only the original \texttt{pnjlim}, \texttt{fetlim} and \texttt{limvds}.
The \texttt{sinhlim} we have developed for RRAM models, is a new one.
Moreover, among all these limiting functions, \texttt{sinhlim} is the only one
that is smooth and continuous, making it more robust to use in simulation. 

\section{\normalfont {\large Models for General Memristive Devices}}\seclabel{memristor}

In this section, we apply the modelling techniques and methodology we have
developed in previous sections to the modelling of general memristive devices.
We use the same model template we have demonstrated in \secref{hys}, where
$f_1$ specifies the device's I-V relationship, $f_2$ describes the dynamics of
the internal unknown.
For general memristive devices, there are several equations available for $f_1$
and $f_2$, from existing models such as the linear and non-linear ion drift
models \cite{kvatinsky2011memristor}, Simmons tunnelling barrier model
\cite{pickett2009TiO2}, TEAM/VTEAM model
\cite{kvatinsky2013team,kvatinsky2015VTEAM}, Yakopcic's model
\cite{yakopcic2013memristor,yakopcic2014memristor}, \etc.
In this section, we examine the reason why they do not work well in simulation,
especially in DC analysis.
We first summarize the common issues with the $f_1$ and $f_2$ functions used in
them, then examine the individual problems of each $f_1$/$f_2$ function, and
list our improvements in \tabref{f1} and \tabref{f2}.

As discussed earlier, both $f_1$, the I-V relationship, and $f_2$, the internal
unknown dynamics, are often highly non-linear and asymmetric \wrt{} positive
and negative voltages; available $f_1$ and $f_2$ functions often use
discontinuous and fast-growing components in them, \eg, exponential, sinh
functions, power functions with a large exponent, \etc.
These components result in difficulty of convergence in simulation.
To overcome these difficulties, similar to what we did in \secref{RRAM} for the
RRAM model, we can use smooth and safe functions.

The key idea of the design of smooth functions is to combine common elementary
functions to approximate the original non-smooth ones.
A parameter common to all these functions, \aka{} smoothing factor, is used to
control the trade-off between better approximation and more smoothness, which
is often synonymous to better convergence. 
Similar ideas apply to safe functions.
For the fast-growing functions, their ``safe'' versions limit the maximum slope
the functions can reach, then linearize the functions to keep the slopes
constant beyond those points.
For functions that are not defined for all real inputs, \eg, sqrt, log, \etc,
their ``safe'' versions clip the inputs using smoothclip such that these
functions will never get invalid inputs. 

Specifically, for the available $f_1$ and $f_2$ functions, the
\texttt{if-then-else} statements can be replaced with smoothswitch.
The exp and sinh functions can be replaced with safeexp and safesinh.
The power functions, \eg, pow(a, b), can also be replaced with
safeexp(b*safelog(a)). 

We have implemented common smooth and safe functions in MAPP.
For example, issuing ``help smoothclip'' within MAPP will display more
information on the usage of smoothclip.
For Verilog-A, we have implemented these smooth and safe functions as
``\texttt{analog functions}'', listed them in a separate file in
\appref{smoothfunctions_va_code} for model developers to use conveniently.

The use of smooth and safe functions are more than numerical tricks, and they
do not necessarily make models less physical.
On the contrary, physical systems are usually smooth.
For example, when switching the voltage of a two-terminal device across zero,
the current should change continuously and smoothly.
Therefore, compared with the original \texttt{if-then-else} statements, the
smoothswitch version is likely to be closer to physical reality.
The same applies to the safe functions we use in our models.
For example, there are no perfect exponential relationships in physical
reality.
Even the growth rate of bacteria, which is often characterized as exponential
in time, will saturate eventually.
Another quantity often modelled using exponential functions is the current
through a p-n junction.
When the voltage indeed becomes large, the junction doesn't really give out
next to infinite current. Instead, other factors come into play --- 
the temperature will become too high that the structure will melt.
This is not considered when writing the exponential I-V relationship;
the use of exponential function is not to capture the physics exactly, but more
an approximation and simplification of physical reality.
So the use of safeexp and safesinh is more than just a means to prevent
numerical explosion, but also a fix to the original over-simplified models.

\begin{table}[htb]
{\scriptsize
  \begin{center}
    \begin{tabular}{|m{0.02\linewidth}|m{0.4\linewidth}|m{0.5\linewidth}|}
    \hline
    {\bf No.} &
    {\bf Original $f_1$} &
    {\bf Comments and improved $f_1$} \\ \hline
    \hline
    1 &
    \[ f_1 = (R_{on}\cdot s + R_{off}\cdot (1-s))^{-1} \cdot vpn.  \] &
    Can have division-by-zero when $s=R_{off}/(R_{on}-R_{off})$. \newline
    We use
    \[
    y = \text{smoothclip}(s - R_{off}/(R_{on}-R_{off}),~smoothing) + R_{off}/(R_{on}-R_{off}),
    \]
    then
	\[f_1 = (R_{on}\cdot y + R_{off}\cdot (1-y))^{-1} \cdot vpn.\]
    \\ \hline
    2 &
    \[ f_1 = \frac{1}{R_{on}} \cdot e^{-\lambda \cdot (1-s)} \cdot vpn.  \] &
    We change exponential function to \texttt{safeexp()}.
    \\ \hline
    3 &
    \[f_1 = s^n \cdot \beta \cdot \sinh(\alpha\cdot vpn) + \chi \cdot
     (\exp(\gamma \cdot vpn) - 1).\] &
    We change sinh to \texttt{safesinh()}, exponential function to
    \texttt{safeexp()}.
    \\ \hline
    4 &
    \[ f_1 = 
	\begin{cases}
		A_1 \cdot s \cdot \sinh(B \cdot vpn), & \text{if } vpn\ge 0\\
		A_2 \cdot s \cdot \sinh(B \cdot vpn), & \text{otherwise}.
	\end{cases} \]
    &
    We change sinh to \texttt{safesinh()}, then smooth the function.
    \[f_{1p} = A_1 \cdot s \cdot \text{safesinh}(B \cdot vpn,~maxslope), \]
	\[f_{1n} = A_2 \cdot s \cdot \text{safesinh}(B \cdot vpn,~maxslope), \]
	\[f_1 = \text{smoothswitch}(f_{1n}, f_{1p}, vpn,~smoothing). \]
    \\ \hline
    5 &
    \[f_1 = I_0\cdot \exp\left(-\frac{Gap}{g_0}\right) \cdot \sinh(\frac{vpn}{V_0}).\] & 
	We express $Gap$ using $s$:
	\[ Gap = s\cdot minGap+(1-s)\cdot maxGap.\]
    Then we change sinh to \texttt{safesinh()}, exponential function to
    \texttt{safeexp()}.
    \\ \hline
    \end{tabular}
  \end{center}
  \caption{\normalfont{The available I-V relationships ($f_1(vpn,~s)$
functions) for general memristive devices, their problems and our improvements.
} \tablabel{f1}}
}
\end{table}

One common problem with existing $f_2$ functions is the range of the internal
unknown.
We have discussed this problem in \secref{RRAM} in the context of RRAM device
models.
The $f_2$ functions available either neglect this issue or use window functions
to set the bounds for the internal unknown.
From the discussion in \secref{RRAM}, using window functions introduces
modelling artifacts that limit the usage of the model to only transient
simulation.
To fix this problem, we apply the same modelling technique using clipping
functions in our memristor models.

\begin{table}[!htbp]
{\scriptsize
  \begin{center}
    \begin{tabular}{|m{0.02\linewidth}|m{0.4\linewidth}|m{0.5\linewidth}|}
    \hline
    {\bf No.} &
    {\bf Original $f_2$} &
    {\bf Comments and improved $f_2$} \\ \hline
    \hline
    1 &
	Linear ion drift model:
    \[ f_2 = \mu_v \cdot R_{on} \cdot f_1(vpn,~s). \] \vspace{-1em} &
    No DC hysteresis. Doesn't ensure $0 \leq s \ge 1$. \newline
    We use the clipping technique to set bounds for $s$.
    \\ \hline
    2 &
	Nonlinear ion drift model:
    \[ f_2 = a \cdot vpn^m. \] \vspace{-1em} &
    No DC hysteresis. Doesn't ensure $0 \leq s \ge 1$. \newline
    We use the clipping technique to set bounds for $s$.
    \\ \hline
    3 &
	Simmons tunnelling barrier model:
	{\tiny
    \[ f_2 = 
	\begin{cases}
		c_{off} \cdot \sinh(\frac{i}{i_{off}}) \cdot
        \exp(-\exp(\frac{s-a_{off}}{w_c} - \frac{i}{b}) - \frac{s}{w_c}),
		& \text{if } i \ge 0\\
		c_{on} \cdot \sinh(\frac{i}{i_{on}}) \cdot
		\exp(-\exp(\frac{a_{on}-s}{w_c} + \frac{i}{b}) - \frac{s}{w_c}),
		& \text{otherwise},
	\end{cases} \]
	}
	where $i = f_1(vpn,~s)$.  &
    No DC hysteresis. Doesn't ensure $0 \leq s \ge 1$. Contains fast-growing functions. \newline
    We change sinh to \texttt{safesinh()}, exponential function to \texttt{safeexp()}, then
    implement the smooth version of this \texttt{if-then-else} statement.
    We use the clipping technique to set bounds for $s$.
    \\ \hline
    4 &
	VTEAM model: \vspace{2em}
    \[
    f_2 = 
	\begin{cases}
		k_{off} \cdot (\frac{vpn}{v_{off}}-1)^{\alpha_{off}}, & \text{if } vpn > v_{off}\\
		k_{on} \cdot (\frac{vpn}{v_{on}}-1)^{\alpha_{on}}, & \text{if } vpn < v_{on}\\
		0, & \text{otherwise}
	\end{cases}
    \]
    &
	DC hysteresis is modelled by a $f_2=0$ flat region. We redesign the equation
    based on \figref{memristor_f2}.
    \[
    f_2 = 
	\begin{cases}
		k_{off} \cdot (\frac{vpn - v^*}{v_{off}})^{\alpha_{off}}, & \text{if } vpn > v^*\\
		k_{on} \cdot (\frac{vpn - v^*}{v_{on}})^{\alpha_{on}}, & \text{otherwise},
	\end{cases}
    \]
	where \[v^* = (1-s) \cdot v_{off}+ s \cdot v_{on}, \]
    such that when $s=1$ and $s=0$, it is equivalent to VTEAM equation in the
    $vpn > v_{off}$ and $vpn < v_{on}$ regions respectively. \newline
    We also make the function smooth:
	\[ f_{2p} = k_{off} \cdot (vpn-v^*/v_{off})^{\alpha_{off}}, \]
	\[ f_{2n} = k_{on} \cdot (vpn-v^*/v_{on})^{\alpha_{on}}, \]
	\[ f_2 = \text{smoothswitch}(f_{2n}, f_{2p}, vpn - v^*,~smoothing). \]
	And finally, we use the clipping technique to set bounds for $s$.
    \\ \hline
    5 &
	Yakopcic's model: 
    \[
    f_2 = g(vpn) \cdot f(s),
    \]
	where
    \[
    g(vpn) = 
	\begin{cases}
		Ap  \cdot (\exp(vpn) - \exp(V_p)), & \text{if } vpn > V_p\\
		-An \cdot (\exp(-vpn) - \exp(V_n)), & \text{if } vpn < -V_n\\
		0, & \text{otherwise},
	\end{cases}
    \]
	and
    \[
    f(s) = 
	\begin{cases}
		\exp(-\alpha_p \cdot (s-x_p)), & \text{if } s \ge x_p\\
		\exp(\alpha_n \cdot (s-1+x_n)), & \text{if } s \leq 1-x_n\\
		1, & \text{otherwise}
	\end{cases}
    \] & 
	DC hysteresis is modelled by a $f_2=0$ flat region. We redesign the equation
    based on \figref{memristor_f2}.
    \[
    g(vpn) = 
	\begin{cases}
		Ap  \cdot (\exp(vpn) - \exp(v^*)),
        & \text{if } vpn > v^*\\
		-An \cdot (\exp(-vpn) - \exp(-v^*)), & \text{otherwise},
	\end{cases}
	\]
	where \[v^* = -V_n \cdot s + V_p \cdot (1-s).\]

	We also change exponential function to \texttt{safeexp()}, make the
	function smooth, then use the clipping technique to set bounds for $s$.
    \\ \hline
    6 &
	Standford/ASU RRAM model:
	\[ f_2 = -v_0 \cdot \exp(- \frac{q\cdot E_a}{k\cdot T}) \cdot
	\sinh(\frac{vpn \cdot \gamma\cdot a_0\cdot q}{k\cdot T \cdot t_{ox}}),\]
	where
	\[ \gamma = \gamma_0 - \beta_0 \cdot Gap^3. \] & 
	We convert $d/dt~Gap$ to $d/dt~s$:
	\[ f_2 = (maxGap-minGap) \cdot v_0 \cdot \exp(- \frac{q\cdot E_a}{k\cdot T}) \cdot
	\sinh(\frac{vpn \cdot \gamma\cdot a_0\cdot q}{k\cdot T \cdot t_{ox}}).\]
    Then we change sinh to \texttt{safesinh()}, exponential function to
    \texttt{safeexp()}.
    We also use the clipping technique to set bounds for $s$.
    \\ \hline
    \end{tabular}
  \end{center}
  \caption{\normalfont{Available internal unknown dynamics ($f_2(vpn,~s)$
functions) for memristive devices, their problems and our improvements.
} \tablabel{f2}}
}
\end{table}

Another problem with the available $f_2$ functions is the way they handle DC
hysteresis.
As discussed earlier, DC hysteresis is observed in forward and backward DC
sweeps;
it accounts for the pinched I-V curves when voltage is moving infinitely slow.
From the model example \texttt{hys\_example} in \secref{hys}, we can conclude that DC hysteresis
results from the model's DC solution curve folding backward in voltage, which
creates multiple stable solutions of internal state variable at certain
voltages.
In fact, from the equations of TEAM/VTEAM model and Yakopcic's model, we can see
an attempt to model DC hysteresis.
However, the way it is done in both these models is to set $f_2=0$ within a
certain voltage range, \eg, when voltage is close to 0.
In this way, as long as the voltage is within this range, there are infinitely
many solutions for the model, regardless of values of $s$.
During transient simulation, $s$ will just keep its old value from the previous
time point.
In DC analysis, if $s$ also keeps its old value from the last sweeping point,
there can be DC hysteresis.
However, since $s$ actually has infinitely many solutions within this voltage
range, the equation system becomes ill-conditioned.
The circuit Jacobian matrix can also become singular, since $s$ has no control
over the value of $f_2$.
Homotopy analysis won't work with these device models since there is no
solution curve to track.
Even in DC operating point (OP) analysis, the OP can have a random $s$ as part
of the solution, depending on the initial condition, and if it is not provided,
on how the OP analysis is implemented.
DC sweep results also depend on how DC sweep is written, particularly on the
way the old values are used as initial guesses for current steps.\footnote{For
example, if the DC sweep implements predictor, when sweeping
across the hysteresis range of voltage, $s$ may not stay flat.}
In other words, because of the model is ill-conditioned, the behaviour of the
model is specific to the implementation of the analysis and will vary from
simulator to simulator.
To fix this problem, we modify the available $f_2$ functions such that the
$f_2=0$ solutions form a single curve in state space, as illustrated in
\figref{memristor_f2} (b).
For each model, this requires different modifications specific to its
equations; we list more detailed descriptions of these modifications in
\tabref{f2}.

\begin{figure}[htbp]
\centering{
    \epsfig{file=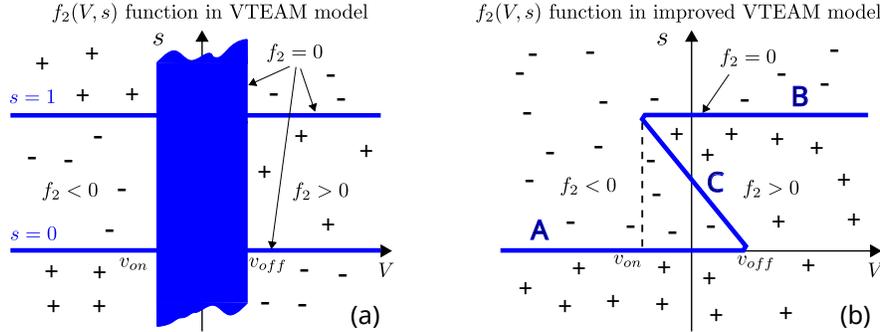,width=0.7\linewidth}
}
\caption{$f_2$ function in VTEAM memristor model contains a flat region around
$V=0$ for the modelling of DC hysteresis. The proper way is to design a single
solution curve of $f_2=0$ that folds back around $V=0$, just like the $f_2$ of
\texttt{hys\_example} in \secref{hys}.
    \figlabel{memristor_f2}}
\end{figure}

To summarize the problems with existing memristor models and our solutions to
them, we fix the nonsmoothness and overflow problems of the existing equations
with smooth and safe functions;
we fix the internal state boundry problem with the same clipping function
technique we have used for the RRAM model;
we fix the ``flat'' $f_2$ problem by properly implementing the $f_2=0$ curve
that bends backward for the modelling of DC hysteresis.
\tabref{f1} and \tabref{f2} list our approaches in improving the available
$f_1$ and $f_2$ functions in more detail.
The result is a collection of memristor models, controlled by two variables
(which can be thought of as higher-level model parameters), f1\_switch and
f2\_switch.
All the combinations of 5 $f_1$ functions and 6 $f_2$ functions constitute 30
compact models for various types of memristors.
Different $f_1$ and $f_2$ functions describe different underlying physics of
the devices, with different levels of accuracy.
We would like to note that one particular combination --- f1\_switch = 5,
f2\_switch = 6, is equivalent to the RRAM model we have discussed in
\secref{RRAM}.

Apart from this combination for RRAM devices, several other combinations in the
general memristor model can also be used for RRAM devices.
For example, when f2\_switch =5 and f2\_switch = 4, our proposed model uses the
improved equations from the VTEAM and Yakopcic's models.
The range of the DC hysteresis in these models is controlled by two threshold
voltages, \eg, $V_p$ and $V_n$ for Yakopcic's model, $v_{off}$ and $v_{on}$ for
VTEAM model.
When both these two thresholds are equal to zero, the DC hysteresis disappears,
and the models are suitable for RRAM devices.
Also, when the two threshold voltages have the same sign, these models can also
be used for unipolar memristive devices.
They are more general and flexible than the model equations we have discussed
in \secref{RRAM} written only for bipolar RRAM devices.
The ideas and techniques underlying these models are likely to also be
applicable to new memristive devices and model equations to be developed in the
future.

\begin{figure}[htbp]
    \begin{minipage}{0.18\linewidth}
      \epsfig{file=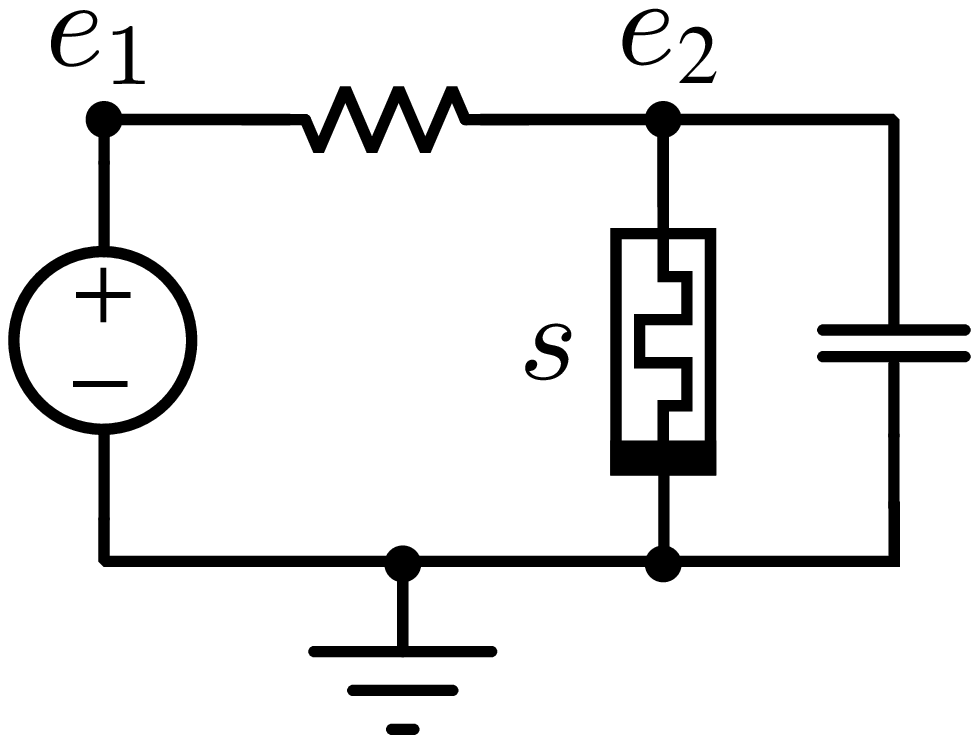,width=\linewidth}
      \caption{Schematic of an oscillator made with unipolar RRAM device.
       }\figlabel{memristor_osc}
    \end{minipage}
    \hfill
    \begin{minipage}{0.75\linewidth}
      \epsfig{file=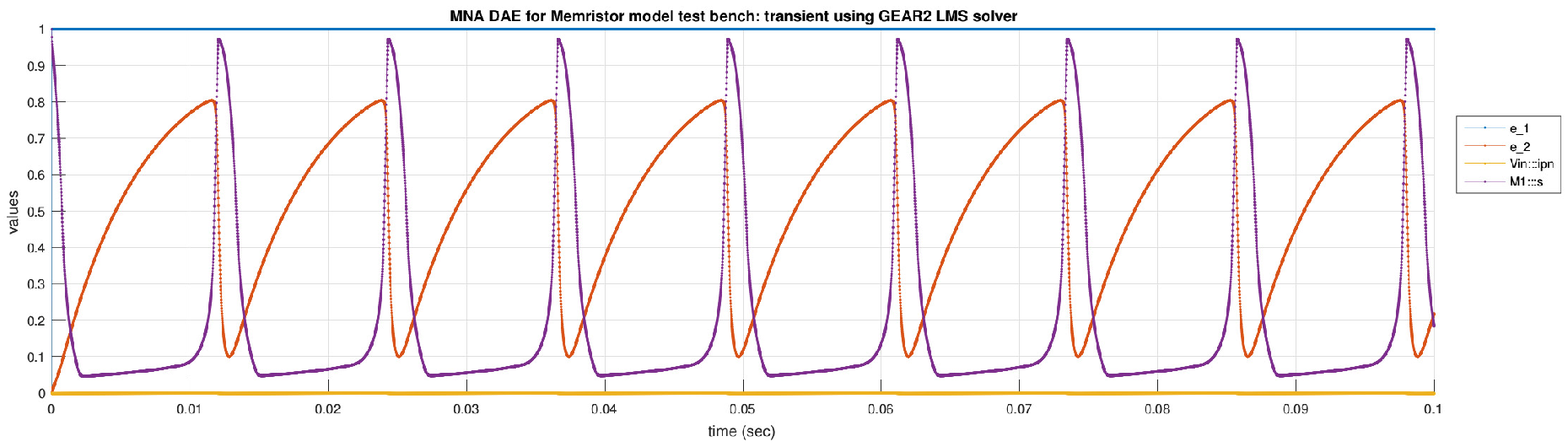,width=\linewidth}
      \caption{Transient simulation results of the RRAM oscillator in \figref{memristor_osc}.
       }\figlabel{memristor_osc_tran}
    \end{minipage}
\end{figure}

\begin{figure}[htbp]
\centering{
    \epsfig{file=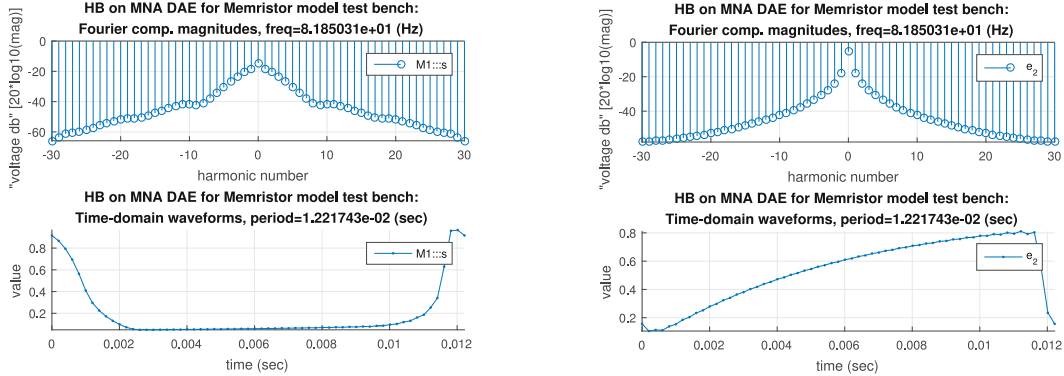,width=0.9\linewidth}
}
\caption{Frequency- and time-domain Harmonic balance results of $s$ and $e_2$
         in the RRAM oscillator.
    \figlabel{memristor_HB}}
\end{figure}

The ModSpec and Verilog-A files of the proposed general memristor models are
listed in \appref{memristor_ModSpec_code} and \appref{memristor_va_code}
respectively.
They can be used in the same test benches for RRAMs in \secref{RRAM}.
Their parameters can also be fitted to generate similar results in
\figref{RRAM_v0_tran} and \figref{RRAM_v0_homotopy}.
As an extra example, we use f1\_switch=2, f2\_switch=5, corresponding to the
improved Yakopcic model, and adjust its parameters for a unipolar RRAM device,
connect it with a resistor as shown in \figref{memristor_osc} to make an
oscillator.
Then we run both transient simulation and PSS analysis with Harmonic Balance
and show their results in \figref{memristor_osc_tran} and
\figref{memristor_HB}.
These results demonstrate that our model not only run in DC, transient and
homotopy analyses, but also work for PSS simulation.

\section{\normalfont {\large Summary}} \seclabel{conclusion}

Our study in this paper centers around the compact modelling of memristive
devices.
Memristor models available today do not work well in simulation, especially in
DC analysis.
Their problems come from several main sources.
Firstly, some models are not in the differential equation format; they are
essentially hybrid models with memory states used for hysteresis.
We clarified that the proper modelling of hysteresis should be achieved
through the use of an internal state variable and an implicit equation.
To make this concept clear, we developed a model template and implemented an
example, namely \texttt{hys\_example}, in both ModSpec and Verilog-A.
During this process, we examined the common mistakes model developers make when
writing internal unknowns and implicit equations in the Verilog-A language.
Then we applied the model template to model RRAM devices, which led to another
common difficulty in memristor modelling --- enforcing the upper and lower
bounds of the internal unknown.
We proposed numerical techniques with clipping functions that can modify the
filament growth equation such that the bounds are respected in simulation.
We also discussed the physical justification behind our approaches.
Then we demonstrated that the same techniques can be applied to fix the similar
problems with many other existing memristor models.
As a result, we not only developed a suite of 30 memristor models, all tested to
work with many simulation analyses in major simulators, but also took this
process as an opportunity to identify and document many good and bad modelling
practices.
Both the resulting models and the techniques used in developing them should be
valuable to the compact modelling community.

\let\em=\it
\bibliographystyle{unsrt}
{\scriptsize\bibliography{stringdefs,tianshi,jr}}

\appendices

\section{\normalfont Model and Circuit Code for hys --- A Device Example with Hysteresis}
\applabel{hys}

 \subsection{\normalfont \texttt{hys\_ModSpec.m}: model file for hys in MAPP}
 \applabel{hys_ModSpec_code}
 \matlabscript{code/hys_ModSpec.m}{lst:hys_ModSpec}{\texttt{hys\_ModSpec.m}}

 \subsection{\normalfont \texttt{hys.va}: Verilog-A model for hys}
 \applabel{hys_va_code}
 \verilogascript{code/hys.va}{lst:hys_va}{\texttt{hys.va}}

 \subsection{\normalfont \texttt{test\_hys.m}: circuit and test script for hys in MAPP}
 \matlabscript{code/test_hys.m}{lst:test_hys_m}{\texttt{test\_hys.m}}

 \subsection{\normalfont \texttt{test\_hys.cir}: circuit and test script for hys in Xyce}
 \verilogascript{code/test_hys.cir}{lst:test_hys_cir}{\texttt{test\_hys.cir}}

 \subsection{\normalfont \texttt{test\_hys.scs}: circuit and test script for hys in \Spectre}
 \verilogascript{code/test_hys.scs}{lst:test_hys_scs}{\texttt{test\_hys.scs}}

 \subsection{\normalfont \texttt{test\_hys.sp}: circuit and test script for hys in \HSPICE}
 \verilogascript{code/test_hys.sp}{lst:test_hys_sp}{\texttt{test\_hys.sp}}

\section{\normalfont Model and Circuit Code for RRAM version 0}
 \subsection{\normalfont \texttt{RRAM\_v0\_ModSpec.m}: model file for RRAM version 0 in MAPP}
 \applabel{RRAM_v0_ModSpec_code}
 \matlabscript{code/RRAM_v0_ModSpec.m}{lst:RRAM_v0_ModSpec}{\texttt{RRAM\_v0\_ModSpec.m}}

 \subsection{\normalfont \texttt{RRAM\_v0.va}: Verilog-A model for RRAM version 0}
 \applabel{RRAM_v0_va_code}
 \verilogascript{code/RRAM_v0.va}{lst:RRAM_v0_va}{\texttt{RRAM\_v0.va}}

 \subsection{\normalfont \texttt{test\_RRAM\_v0.m}: circuit and test script for RRAM version 0 in MAPP}
 \matlabscript{code/test_RRAM_v0.m}{lst:test_RRAM_v0_m}{\texttt{test\_RRAM\_v0.m}}

 \subsection{\normalfont \texttt{test\_RRAM\_v0.cir}: circuit and test script for RRAM version 0 in Xyce}
 \verilogascript{code/test_RRAM_v0.cir}{lst:test_RRAM_v0_cir}{\texttt{test\_RRAM\_v0.cir}}

 \subsection{\normalfont \texttt{test\_RRAM\_v0.scs}: circuit and test script for RRAM version 0 in \Spectre}
 \verilogascript{code/test_RRAM_v0.scs}{lst:test_RRAM_v0_scs}{\texttt{test\_RRAM\_v0.scs}}

 \subsection{\normalfont \texttt{test\_RRAM\_v0.sp}: circuit and test script for RRAM version 0 in \HSPICE}
 \verilogascript{code/test_RRAM_v0.sp}{lst:test_RRAM_v0_sp}{\texttt{test\_RRAM\_v0.sp}}

\section{\normalfont Model Code for Memristor}
 \subsection{\normalfont \texttt{Memristor.m}: model file for memristor ModSpec model in MAPP}
 \applabel{memristor_ModSpec_code}
 \matlabscript{code/Memristor.m}{lst:Memristor}{\texttt{Memristor.m}}

 \subsection{\normalfont \texttt{Memristor.va}: Verilog-A model for Memristor}
 \applabel{memristor_va_code}
 \verilogascript{code/Memristor.va}{lst:Memristor_va}{\texttt{Memristor.va}}

 \subsection{\normalfont \texttt{smoothfunctions.va}: Verilog-A file for smoothing function definitions}
 \applabel{smoothfunctions_va_code}
 \verilogascript{code/smoothfunctions.va}{lst:smoothfunctions}{\texttt{smoothfunctions.va}}

\end{document}